\documentclass[12pt]{article}
\usepackage[margin=0.75in]{geometry}
\usepackage{xcolor}
\usepackage{amsmath}
\usepackage{amssymb}
\usepackage{bm}
\usepackage{graphicx}
\usepackage{placeins}
\usepackage{caption}
\usepackage{subcaption}
\usepackage{indentfirst}
\usepackage{booktabs}
\usepackage{multirow}
\usepackage{accents}

\newlength{\dhatheight}
\newcommand{\doublehat}[1]{
    \settoheight{\dhatheight}{\ensuremath{\hat{#1}}}
    \addtolength{\dhatheight}{-0.35ex}
    \hat{\vphantom{\rule{1pt}{\dhatheight}}
    \smash{\hat{#1}}}
}
\newcommand{\doublehatsubscript}[1]{
    \settoheight{\dhatheight}{\ensuremath{\hat{#1}}}
    \addtolength{\dhatheight}{-0.9ex}
    \hat{\vphantom{\rule{1pt}{\dhatheight}}
    \smash{\hat{#1}}}
}

\usepackage[numbib,notlof,notlot,nottoc]{tocbibind}

\usepackage{hyperref}
\hypersetup{colorlinks=true,linkcolor=blue,filecolor=magenta,urlcolor=blue}
\newcommand{\mut}{{\tilde{\mu}}}
\newcommand{\mup}{{\mu^\prime}}
\newcommand{\muh}{\hat{\mu}}
\newcommand{\sh}{\hat{s}}
\newcommand{\bh}{\hat{b}}
\newcommand{\bhh}{\doublehat{b}}
\newcommand{\br}[1]{\left(#1\right)}

\newcommand{\cbr}[1]{\left\{#1\right\}}
\newcommand{\prob}[2]{P\br{#1\,\middle|\,#2}}
\newcommand{\pois}[2]{\frac{\br{#2}^{#1}\cdot\exp\br{-\br{#2}}}{#1!}}
\newcommand{\bvec}{\vec{b}}
\newcommand{\bhvec}{\vec{\bh}}

\newcommand{\sumb}{\sum_{i=1}^{N}b_{i}}
\newcommand{\sumbh}{\sum_{i=1}^{N}\bh_{i}}
\newcommand{\sumbhh}{\sum_{i=1}^{N}\bhh_{i}}
\newcommand{\sumtaub}{\sum_{j=1}^{N}\tau_{ij}b_{j}}
\newcommand{\sumtaubh}{\sum_{j=1}^{N}\tau_{ij}\bh_{j}}
\newcommand{\sumtaubhh}{\sum_{j=1}^{N}\tau_{ij}\bhh_{j}}
\newcommand{\pd}[2]{\frac{\partial #1}{\partial #2}}
\newcommand{\that}{\hat{\theta}}
\newcommand{\thh}{\doublehat{\theta}}
\newcommand{\tvec}{\vec{\theta}}
\newcommand{\thvec}{\vec{\that}}
\newcommand{\thhvec}{\vec{\thh}}
\newcommand*{\xhat}[2]{#2\kern+#1cm\hat{\phantom{#2}}}

\usepackage{authblk}
\usepackage{blindtext}
\usepackage[title]{appendix}

\begin{document}

\title{Generalized asymptotic formulae for estimating statistical significance in high energy physics analyses}
\author[1]{M. J. Basso\footnote{E-mail: \href{mailto:mbasso@physics.utoronto.ca}{mbasso@physics.utoronto.ca}}}
\affil[1]{Department of Physics, University of Toronto, 60 St. George St., Toronto, Ontario, Canada}
\date{\today}
\maketitle

\begin{abstract}
Within the framework of likelihood-based statistical tests for high energy physics measurements, we derive generalized expressions for estimating the statistical significance of discovery using the asymptotic approximations of Wilks and Wald for a variety of measurement models. These models include arbitrary numbers of signal regions, control regions, and Gaussian constraints. We extend our expressions to use the representative or ``Asimov'' dataset proposed by Cowan \textit{et al.} such that they are made data-free. While many of the generalized expressions are complicated and often involve solving systems of coupled, multivariate equations, we show these expressions reduce to closed-form results under simplifying assumptions. We also validate the predicted significance using toy-based data in select cases.
\end{abstract}

\section{Introduction}
\label{sec:intro}
\subsection{Relevant Theory}
\label{sec:theory}

In the field of high energy physics (HEP), likelihood-based statistical tests entail the construction of a likelihood function $L$ describing a particular measurement model; the likelihood function in turn describes the ``likelihood'' of measuring parameters defining the model given some observed data~\cite{Cranmer2014}. For counting experiments typical of HEP analyses at the Large Hadron Collider (LHC), the likelihood may be written most simply as a product of Poisson probability density functions (PDFs) over $N$ ``regions'' or ``bins'':

\begin{equation}
    L(\tvec) = \prod_{i=1}^{N}P(n_i|\nu_i(\tvec)) = \prod_{i=1}^{N}\frac{\nu_i(\tvec)^{n_i}\cdot\exp{(-\nu_i(\tvec)})}{n_i!} \,,
\end{equation}

\noindent where $n_i$ and $\nu_i$ are the observed and expected yields in region $i$, respectively, and $\tvec$ are the free parameters defining out model. Here, we have assumed a uniform prior $\pi(\tvec)$ for our free parameters (i.e., no prior knowledge). The best-fit parameters for a given measurement will be those which \textit{maximize} the likelihood.

Typically, one is interested in measuring some signal $s$ (e.g., the number of Higgs boson decay events to $WW^*$) given some known or constrained background $b$ (e.g., the number of Drell-Yan events). In this case, $s$ defines our parameter-of-interest (POI), what we are interested in measuring, while $b$ defines a nuisance parameter (NP), a parameter we measure but which may not be physically interesting. We may parametrize the expected yield as $\nu(\mu, b) = \mu s + b$, where $s$ is now fixed and our signal strength $\mu$ is what tunes the amount of signal, now becoming our POI\footnote{N.B.: it is equally valid to let $s$ be our POI, but in the spirit of consistency with the literature on this topic, we adopt this reparametrization.}. Absorbing $b$ into $\tvec$, which we now assumes contains only our NPs, and letting $L = L(\mu, \tvec)$, we construct the log-likelihood ratio:

\begin{equation}
    \label{eqn:lambda}
    \lambda(\mut) = \frac{L(\mut, \thhvec)}{L(\muh, \thvec)} \,,
\end{equation}

\noindent where $\muh$ and $\thvec$ are the \textit{unconditional} maximum likelihood estimators (MLEs) of $L$ (i.e., the values of $\mu$ and $\tvec$ which set $\partial L / \partial\mu = 0$ and $\partial L / \partial\theta_i = 0 \,\forall\, i = 1,\ldots,N$ where $N$ is the number of NPs) and $\thhvec$ is the \textit{conditional} MLE of $L$ for fixed $\mu = \mut$ (i.e., the values of $\tvec$ which set $\partial L / \partial\theta_i = 0 \,\forall\, i = 1,\ldots,N$ where $N$ is the number of NPs)~\cite{Cranmer2014,Cowan2011} -- $\mut$ is herein referred to as the \textit{hypothesized} value of $\mu$. Eq.~\ref{eqn:lambda} ranges between 0 and 1, where $\lambda(\mut) \sim 1$ indicates good agreement between the hypothesized value of $\mu$ and its MLE while $\lambda(\mut) \ll 1$ indicates disagreement.

Using Eq.~\ref{eqn:lambda}, we define our test statistic:

\begin{equation}
    t_\mut = -2 \ln \lambda(\mut) \,,
\end{equation}

\noindent where $t_\mut \sim 0$ indicates good agreement between the hypothesized $\mu$ and its MLE and increasing $t_\mut$ indicates increasing disagreement~\cite{Cranmer2014,Cowan2011}. We may define a $p$-value, $p_\mut$, representing the probability of observering equal or greater disagreement with the hypothesized $\mu$ as:

\begin{equation}
    \label{eqn:t_mu}
    p_\mut = \int_{t_{\mut,\textrm{obs}}}^{\infty} f(t_{\mut}|\mut)\,dt_{\mut} \,,
\end{equation}

\noindent where $f(t_{\mut}|\mut)$ is the PDF of the test statistic assuming hypothesized $\mu = \mut$ and $t_{\mut,\textrm{obs}}$ is the observed value of the test statistic. As is common practice in HEP, one may translate the $p$-value into the number of standard deviations from the mean of a standard Gaussian whose integrated, one-sided tail equals such a probability, i.e.:

\begin{equation}
    \label{eqn:Z_mu}
    Z_\mut = \Phi^{-1}(1-p_\mut) \,,
\end{equation}

\noindent where $\Phi^{-1}$ is the inverse cumulative function of a standard Gaussian. This quantity is referred to as the \textit{significance} or the \textit{sensitivity} of the measurement.

In ``discovery'' HEP analyses, one looks to measure the presence of signal (i.e., $\mu s + b$ with $\mu > 0$) among background processes, adopting the null hypothesis $H_0$ that no signal is present (i.e., $\mut = 0$) and the alternative hypothesis $H_1$ that signal is present in some fixed amount. An example of a discovery analysis would be a measurement of vector boson fusion production of Higgs bosons decaying to $WW^*$ over the prevailing top quark pair and single top production, Drell-Yan, and diboson backgrounds.

Within the framework of likelihood-based statistical tests for HEP, we adopt the test statistic for discovery analyses proposed by Cowan \textit{et al.}~\cite{Cowan2011}:

\begin{equation}
    \label{eqn:disc_t0}
    t_0 = \begin{cases} -2 \ln \lambda(0) & \text{if $\muh \geq 0$} \\  0 & \text{if $\muh < 0$} \\ \end{cases} \,,
\end{equation}

\noindent where, as before, $\muh$ is our unconditional MLE of $\mu$. As we should not measure a negative signal strength for a signal model predicting an enhancement to our measured yields, we set $t_0$ equal to 0 as a lower bound on our test statistic (i.e., consistent with the null hypothesis). By Eq.~\ref{eqn:t_mu}, our $p$-value becomes:

\begin{equation}
    \label{eqn:t_0}
    p_0 = \int_{t_{0,\textrm{obs}}}^{\infty} f(t_0|0)\,dt_0 \,,
\end{equation}

\noindent and by Eq.~\ref{eqn:Z_mu}, our significance of discovery is:

\begin{equation}
    \label{eqn:Z_0}
    Z_0 = \Phi^{-1}(1-p_0) \,.
\end{equation}

\noindent To claim the discovery of a signal, it is typical to require that the significance exceeds $5\sigma$: $Z_0 \geq 5$. This corresponds to exluding the null hypothesis at the level of $p_0 = 2.87 \cdot 10^{-7}$.

Often, a physicist will want to know the expected significance of a measurement assuming their signal model in MC to be true and correct. In the case of discovery analyses, this will necessitate knowledge of $f(t_0|0)$, the PDF of the test statistic assuming no signal. An approximation of $t_{0,\textrm{obs}}$ may be made by setting it equal to the median value of $t_0$ distributed according to $f(t_0|\mup)$, the PDF of the test statistic for discovery assuming a \textit{true} signal strength $\mup$. As an equation, the median $p$-value assuming a true signal strength $\mup$ is given by:

\begin{equation}
    \label{eqn:med_p0}
    \textrm{med}[p_0|\mup] = \int_{\textrm{med}[t_0|\mup]}^{\infty} f(t_0|0)dt_0 \,.
\end{equation}

Without knowing $f(t_0|0)$ and $f(t_0|\mup)$, the above expression is difficult to evaluate. Using the approximations of Wilks~\cite{Wilks1938} and Wald~\cite{Wald1943}, Cowan \textit{et al.}~\cite{Cowan2011} show that $p$-value for discovery may be approximated as:

\begin{equation}
    \label{eqn:p0_asymp}
    p_0 = 1 - F(t_0|0) \approx 1 - \Phi(\sqrt{t_0}) \,,
\end{equation}

\noindent where $F(t_0|0)$ is the cumulative distribution function (CDF) for $f(t_0|0)$. The approximation is valid in the asymptotic limit (i.e., $1/\sqrt{N} \ll 1$ where $N$ is the sample size) and assuming the best-fit signal strength $\muh$ is Gaussian distributed. Inserting Eq~\ref{eqn:p0_asymp} into Eq.~\ref{eqn:Z_0} yields:

\begin{equation}
    \label{eqn:z0_asymp}
    Z_0 \approx \sqrt{t_0} \,,
\end{equation}

\noindent under the same assumptions. Given that $Z_0$ is a monotonically decreasing function of $p_0$ and using Eqs.~\ref{eqn:Z_0}, \ref{eqn:med_p0}, and \ref{eqn:z0_asymp}, we may also write:

\begin{equation}
    \label{eqn:med_z0_asymp}
    \textrm{med}[Z_0|\mup] = \Phi^{-1}\!\br{1 - \int_{\textrm{med}[t_0|\mup]}^{\infty} f(t_0|0)dt_0} \approx \sqrt{\textrm{med}[t_0|\mup]} \,.
\end{equation}

To evaluate the above, Cowan \textit{et al.} propose the use of the ``Asimov'' dataset where the estimators of all parameters yield their true values~\cite{Cowan2011}. In our above formulae, this is equivalent to setting all parameters equal to their true values given our particular physics model (e.g., $\muh \rightarrow \mup$ and $n \rightarrow \mup s + b$). If $t_{0,\textrm{A}}$ is the Asimov value of our test statistic for discovery assuming the true signal strength $\mup$, then we can write:

\begin{equation}
    \label{eqn:asimov_t0}
    \textrm{med}[t_0|\mup] = t_{0,\textrm{A}} \,,
\end{equation}

\noindent and by inserting Eq.~\ref{eqn:asimov_t0} into Eq.~\ref{eqn:med_z0_asymp}, we yield:

\begin{equation}
    \label{eqn:asimov_t0_2}
    \textrm{med}[Z_0|\mup] \approx \sqrt{t_{0,\textrm{A}}} \,.
\end{equation}

\noindent This is one of the important results shown by Cowan \textit{et al.}~\cite{Cowan2011}. It says we can estimate the median significance of discovery as the square root of the test statistic for discovery evaluted using Asimov data. Using the above, one can produce analytical approximations for a variety of measurement scenarios, giving a physicist a handle on the expected power of their analysis techniques without relying on numerical recourse. As Cowan \textit{el al.} discuss in their paper, the asymptotic approximation is already quite good for $N \sim \mathcal{O}(100)$ (see for instance Fig.~7 of Ref.~\cite{Cowan2011}).

This note proceeds as follows: we will motivate and construct several different measurement scenarios (e.g., multiple control regions, multiple signal regions, etc.) a physicist typically encounters and derive expressions for the median significance of discovery using the asymptotic approximation and assuming Asimov data. In all cases, we generalize to an arbitrary number of regions or constraints $N$ and show that the resulting formulae reduce to expected formulae (i.e., derived elsewhere) in the $N = 1$ case or to agree with numerical simulation in test cases.

Additionally, we will simplify the use of Eq.~\ref{eqn:asimov_t0_2} in the following sections by dropping the approximation (i.e., setting it to an equality) and by assuming $\mup = 1$, typical of discovery analyses where the signal model's cross section is normalized to theoretical expectations. We define:

\begin{equation}
    \label{eqn:Z0_simple}
    Z_0 \equiv \textrm{med}[Z_0|\mup = 1] = \sqrt{t_{0,\textrm{A}}} = \sqrt{-2 \ln\!\br{\frac{L(0, \thhvec_\textrm{A})}{L(\muh_\textrm{A}, \thvec_\textrm{A})}}}\,,
\end{equation}

\noindent where we have inserted Eq.~\ref{eqn:disc_t0} followed by Eq.~\ref{eqn:lambda} and the best-fit values of $\thhvec$, $\muh$, and $\thvec$ are assumed to be evaluated using Asimov data (hence the subscript ``A'').

\subsection{Numerical Simulation}
\label{sec:simulation}

To verify our derivations for measurement scenarios which have not yet been studied analytically, we will draw toy events from the PDF governing the measurement scenario at hand in each of the relevant regions and with the PDF's NPs set to their true values. Using the Python package \texttt{probfit}~\cite{probfit} to set up a simultaneous, unbinned (in each region), maximum-likelihood fit and MINUIT~\cite{minuit} via the Python package \texttt{iminuit}~\cite{iminuit} to perform the minimization, we will extract the minima of $-2\ln L(0, \thh)$ and $-2\ln L(\muh, \that)$, allowing us to calculate our test statistic $t_0$ using Eqs.~\ref{eqn:lambda}~and~\ref{eqn:disc_t0}. By performing this procedure many (i.e., $\mathcal{O}(10000)$) times assuming $\mup = 0$ and then assuming $\mup = 1$, we can produce approximate PDFs of the test statistic, $f(t_0|\mup=0)$ and $f(t_0|\mup=1)$. By integrating $f(t_0|\mup=0)$ from the median value of $f(t_0|\mup=1)$ to infinity, we yield the $p$-value of the measurement which then yields the median significance of discovery using Eq.~\ref{eqn:med_z0_asymp}. The Python packages \texttt{numpy}~\cite{numpy}, \texttt{scipy}~\cite{scipy}, and \texttt{matplotlib}~\cite{matplotlib} are used for processing and plotting.

The code implementing the asymptotic formulae and simulations described in this paper is publically available in the following Git respository~\cite{Basso2021}:

\begin{center}
    \texttt{\url{https://github.com/mjbasso/asymptotic_formulae_examples}} \,.
\end{center}

\noindent The respository also includes scripts for producing all of the plots included in this paper.

\FloatBarrier

\section{Derivations}
\label{sec:derivations}
In following subsections, we will derive expressions for the median significance of discovery in the asymptotic limit for a variety of commonly encountered measurement scenarios and provide validation of some of our expressions by comparisons to other sources or by numerical simulation. In particular, we will cover:

\begin{itemize}
    \item Section~\ref{sec:1SRNCR}: 1 signal region + $N$ control regions, $N \in \mathbb{N}$;
    \item Section~\ref{sec:NSR1CR}: $N$ signal regions + 1 control region, $N \in \mathbb{N}$;
    \item Section~\ref{sec:NSRMCR}: $N$ signal regions + $M$ control regions, $N,M \in \mathbb{N}$;
    \item Section~\ref{sec:gauss}: 1 signal region containing $N$ background processes with $M$ Gaussian background constraints, $N,M \in \mathbb{N}$.
\end{itemize}

\subsection{1 Signal Region + $N$ Control Regions}
\label{sec:1SRNCR}
\subsubsection{General Case}
\label{sec:1SRNCR_general}

Assuming a uniform prior, we can write our likelihood with 1 signal region (SR) and $N$ orthogonal auxiliary measurements (read as: $N$ control regions (CRs) for backgrounds $b_{i}$, $i=1,{\ldots},N$) as:

\begin{equation}
    L(s,\vec{b}) = \prob{n}{s + \sum_{i=1}^{N} b_{i}} \cdot \prod_{i=1}^{N} \prob{m_{i}}{\sumtaub} \,,
\end{equation}

\noindent where $P$ refers to a Poisson PDF, $n$ is the observed yield in our SR, $m_{i}$ is the observed yield in CR~$i$, and $\tau_{ij}$ are the transfer factors which carry background $j$ in our SR to CR~$i$. Inserting the mathematical form for $P$ yields:

\begin{equation}
    L(s,\vec{b}) = \pois{n}{s+\sumb} \cdot \prod_{i=1}^{N} \pois{m_{i}}{\sumtaub} \,,
\end{equation}

\noindent and taking the logarithm yields:

\begin{equation}
    \ln{L(s,\vec{b})} = n \cdot \ln\br{s+\sumb} - s - \sumb + \sum_{i=1}^{N}\br{ m_{i} \cdot \ln\br{\sumtaub} - \sumtaub} \,.
\end{equation}

\noindent As we are dealing with a likelihood, constant offsets do not affect our optimization and so we have dropped $-(\ln(n!)+\sum_{i=1}^{N}\ln(m_{i}!))$.

We first consider the most probable value for the backgrounds, $\bhh_{i}$, in the absence of signal, $s=0$. We are interested in maximizing $\ln{L}$. As signal is fixed and constant, we make this explicit in $\ln{L}$ by evaluating it at $s=0$ \textit{prior} to taking any partial derivatives. The result is then differentiated with respect to background $k$ and evaluated at $\vec{b}=\vec{\bhh}$ to yield:

\begin{equation}
    \pd{\ln L(0,\vec{b})}{b_{k}}\bigg|_{\vec{b}=\vec{\bhh}} = \frac{n}{\sumbhh} - 1 + \sum_{i=1}^{N} \tau_{ik} \cdot \br{ \frac{m_{i}}{\sumtaubhh} - 1} \,.
\end{equation}

\noindent Setting all of the partial derivatives, $k=1,{\ldots},N$, equal to 0 yields the following system of equations for $\vec{\bhh}$:

\begin{equation}
    \label{eqn:ncr_bhh}
    \cbr{0 = \frac{n}{\sumbhh} - 1 + \sum_{i=1}^{N} \tau_{ik} \cdot \br{ \frac{m_{i}}{\sumtaubhh} - 1} \,;\, k=1,{\ldots},N } \,.
\end{equation}

We now consider maximizing the likelihood in the presence of signal $s$. In this situation, we let $\sh$ and $\vec{\bh}$ be the signal and background yields, respectively, which maximize our likelihood. Signal and background yields are both left floating, and so we take the partial derivative with respect to $s$ evaluated at $(s,\vec{b})=(\sh,\vec{\bh})$:

\begin{equation}
    \label{eqn:ncr_dLdsh}
    \pd{\ln L(s,\vec{b})}{s}\bigg|_{(s,\vec{b})=(\sh,\vec{\bh})} = \frac{n}{\sh+\sumbh} - 1 \,,
\end{equation}

\noindent as well as the partial derivative with respect to $b_k$:

\begin{equation}
    \label{eqn:ncr_dLdbh}
    \pd{\ln L(s,\vec{b})}{b_k}\bigg|_{(s,\vec{b})=(\sh,\vec{\bh})} = \frac{n}{\sh+\sumbh} - 1 + \sum_{i=1}^{N} \tau_{ik} \cdot \br{ \frac{m_{i}}{\sumtaubh} - 1} \,.
\end{equation}

\noindent Setting Eqs.~\ref{eqn:ncr_dLdsh} and \ref{eqn:ncr_dLdbh} equal to 0 and substituting Eq.~\ref{eqn:ncr_dLdsh} into Eq.~\ref{eqn:ncr_dLdbh} yields:

\begin{equation}
    0 = \sum_{i=1}^{N} \tau_{ik} \cdot \br{ \frac{m_{i}}{\sumtaubh} - 1} \,.
\end{equation}

\noindent Our system of equations for $\sh$ and $\bhvec$ is then:

\begin{equation}
    \label{eqn:ncr_sol}
    \cbr{\sh = n - \sumbh \,,\, 0 = \sum_{i=1}^{N} \tau_{ik} \cdot \br{ \frac{m_{i}}{\sumtaubh} - 1} \,;\, k=1,{\ldots},N } \,.
\end{equation}

\noindent We make the intuitive ansatz that the solutions to Eq.~\ref{eqn:ncr_sol} are $\sh = s$ and $\bhvec = \bvec$ when assuming Asimov data (i.e., $n = s + \sumb$ and $m_i = \sumtaub \,\forall\, i=1,\ldots,N$). Indeed, this can be explicitly checked:

\begin{equation}
    \begin{split}
        & \sh - n + \sumbh = s - \br{s + \sum_{i=1}^N b_i} + \sum_{i=1}^N b_i = 0 \,, \\
        & \sum_{i=1}^{N} \tau_{ik} \cdot \br{ \frac{m_{i}}{\sumtaubh} - 1} = \sum_{i=1}^{N} \tau_{ik} \cdot \br{ \frac{\sumtaub}{\sumtaub} - 1} = \sum_{i=1}^{N} \tau_{ik} \cdot \br{ 1 - 1} = 0 \,.
    \end{split}
\end{equation}

Using Eq.~\ref{eqn:Z0_simple} and taking $\sh = s$ and $\bhvec = \bvec$, our significance of discovery is:

\begin{equation}
    Z_0 = \sqrt{-2\cdot\br{ n \cdot \ln\!\br{\frac{\sumbhh}{s + \sumb}} + s + \sum_{i=1}^{N}\br{(b_{i} - \bhh_{i}) + m_{i} \cdot \ln\br{\frac{\sumtaubhh}{\sumtaub}} + \sum_{j=1}^{N}\tau_{ij}\cdot(b_{j} - \bhh_{j})} }} \,,
\end{equation}

\noindent In this expression and the expressions for $\bhh_i$, we set $n = s+\sumb$ and $\vec{m} = \sumtaub$. We may simplify the above further by using $n = s + \sumb$ to yield:

\begin{equation}
    \label{eqn:ncr_z0}
    Z_0 = \sqrt{-2\cdot\br{ n \cdot \ln\!\br{\frac{\sumbhh}{n}} + n + \sum_{i=1}^{N}\br{-\bhh_{i} + m_{i} \cdot \ln\br{\frac{\sumtaubhh}{\sumtaub}} + \sum_{j=1}^{N}\tau_{ij}\cdot(b_{j} - \bhh_{j})} }} \,.
\end{equation}

\noindent This is our expression for the median significance of discovery in the asymptotic limit.

\subsubsection{Assuming $N=1$ Control Regions}
\label{sec:1SRNCR_Neq1}

As a check, in the case where we have only 1 CR, $N=1$, we let $\bhh \equiv \bhh_{1}$, $m \equiv m_{1}$, and $\tau \equiv \tau_{11}$. From Eq.~\ref{eqn:ncr_bhh}, we yield:

\begin{equation}
    \label{eqn:1cr_simple_bhh}
    0 = \frac{n}{\bhh} - 1 + \tau \cdot \br{\frac{m}{\tau\cdot\bhh} - 1} \Leftrightarrow \bhh = \frac{n+m}{1+\tau} \,,
\end{equation}

\noindent as expected. Additionally, letting $b \equiv b_{1}$, Eq.~\ref{eqn:ncr_sol} yields:

\begin{equation}
    \label{eqn:1cr_simple_bhsh}
    0 = \tau\cdot\br{\frac{m}{\tau \cdot b} - 1} \Leftrightarrow b = \frac{m}{\tau} \Rightarrow s = n - \frac{m}{\tau} \,,
\end{equation}

\noindent as expected. Finally, from Eq.~\ref{eqn:ncr_z0}, our significance is:

\begin{equation}
    Z_0 = \sqrt{-2\cdot\br{ n \cdot \ln\!\br{\frac{\bhh}{n}} + n - \bhh + m \cdot\ln\!\br{\frac{\bhh}{b}} + \tau\cdot(b-\bhh)}} \,,
\end{equation}

\noindent but we know $\tau \cdot b - (1+\tau)\cdot\bhh = m - (n+m) = -n$ by Eqs.~\ref{eqn:1cr_simple_bhh} and \ref{eqn:1cr_simple_bhsh}, leaving us with:

\begin{equation}
    \label{eqn:1cr_simple_z0}
    \begin{split}
        Z_0 & = \sqrt{-2\cdot\br{ n \cdot \ln\!\br{\frac{\bhh}{n}} + m \cdot\ln\!\br{\frac{\bhh}{b}}}} \\ & = \sqrt{-2\cdot\ln\!\br{\br{\frac{n+m}{1+\tau}}^{n+m}\cdot\frac{\tau^{m}}{n^{n}m^{m}}}} \,,
    \end{split}
\end{equation}

\noindent matching what is shown in Eqs.~21 and 22 of Ref.~\cite{Buttinger2019}

\subsubsection{Assuming Diagonal \texorpdfstring{$\boldsymbol{\tau}$}{tau}}
\label{sec:1SRNCR_diag_tau}

Often CRs are defined such that they yield high-stats, pure regions for a specific background. Here, we assume CR $i$ targets background $i$ by assuming the matrix of transfer factors $\boldsymbol{\tau}$ is diagonal (i.e., the acceptance of CR $i$ is 1 for background $i$ and 0 for all other backgrounds). Letting $\tau_{k} \equiv \tau_{kk}$, our equation for $\vec{\bhh}$, Eq.~\ref{eqn:ncr_bhh}, simplifies as:

\begin{equation}
    \label{eqn:diagonal_tau_bhhk}
    0 = \frac{n}{\sumbhh} - 1 + \frac{m_{k}}{\bhh_{k}} - \tau_{k} \Leftrightarrow \frac{n}{\sumbhh} + \frac{m_{k}}{\bhh_{k}} = 1 + \tau_{k}\,,
\end{equation}

\noindent for $k=1,\ldots,N$. Our significance of discovery is:

\begin{equation}
    Z_0 = \sqrt{-2\cdot\br{ n \cdot \ln\!\br{\frac{\sumbhh}{n}} + n + \sum_{i=1}^{N}\br{-\bhh_{i} + m_{i} \cdot \ln\br{\frac{\bhh_{i}}{b_{i}}} + \tau_{i}\cdot(b_{i} - \bhh_{i})} }} \,,
\end{equation}

\noindent where we used $\tau_{ij} = \tau_{i}\cdot\delta_{ij}$, where $\delta_{ij}$ is the Kronecker delta function. Or, given $(-n\cdot\bhh_{k})/(\sumbhh) = m_{k} - (1+\tau_{k})\cdot\bhh_{k} = -\bhh_{k} + \tau_{k}\cdot(b_{k}-\bhh_{k})$ by Eq.~\ref{eqn:diagonal_tau_bhhk}, we can also write:

\begin{equation}
    \label{eqn:ncr_z0_diag}
    \begin{split}
        Z_0 & = \sqrt{-2\cdot\br{ n \cdot \ln\!\br{\frac{\sumbhh}{n}} + n + \sum_{i=1}^{N}\br{m_{i} \cdot \ln\br{\frac{\bhh_{i}}{b_{i}}} - \frac{n\cdot\bhh_{i}}{\sum_{j=1}^{N}\bhh_{j}}} }} \\ & = \sqrt{-2\cdot\ln\!\br{\br{\frac{\sumbhh}{n}}^n\cdot\prod_{i=1}^N\br{\frac{\bhh_{i}}{b_{i}}}^{m_i}}} \,.
    \end{split}
\end{equation}

\subsubsection{Assuming $N=2$ Control Regions and Diagonal \texorpdfstring{$\boldsymbol{\tau}$}{tau}}
\label{sec:1SRNCR_Neq2_diag_tau}

As a special case of the previous section, we consider 2 CRs ($N=2$) and assume each CR to be pure in the background they target, i.e., $\boldsymbol{\tau}$ is diagonal. Then by Eq.~\ref{eqn:diagonal_tau_bhhk}:

\begin{equation}\label{eqn:1cr_system_of_eqns}
    \cbr{ 0 = \frac{n}{\bhh_{1} + \bhh_{2}} - 1 + \br{\frac{m_{1}}{\bhh_{1}} - \tau_{1}} \,,\, 0 = \frac{n}{\bhh_{1} + \bhh_{2}} - 1 + \br{\frac{m_{2}}{\bhh_{2}} - \tau_{2}} } \,,
\end{equation}

\noindent and subtracting the second from the first yields:

\begin{equation}
    \label{eqn:2cr_b1b2}
    0 = \br{\frac{m_{1}}{\bhh_{1}} - \frac{m_{2}}{\bhh_{2}}} - \br{\tau_{1} - \tau_{2}} \Leftrightarrow \bhh_{1} = \frac{m_{1} \cdot \bhh_{2}}{m_{2} + \br{\tau_{1} - \tau_{2}} \cdot \bhh_{2}} \,.
\end{equation}

Consider the simpler case where $\tau_{1} = \tau_{2} = \tau > 0$. Then $\bhh_{1} = \frac{m_{1}}{m_{2}}\cdot\bhh_{2}$ and:

\begin{equation}
    \begin{split}
        0 & = n\cdot\bhh_{2} - (1+\tau)\cdot(\bhh_{1}+\bhh_{2})\cdot\bhh_{2}+m_{2}\cdot(\bhh_{1}+\bhh_{2}) \\
        & = n\cdot\bhh_{2} - (1+\tau)\cdot\frac{m_{1} + m_{2}}{m_{2}}\cdot\bhh_{2}^{2} + (m_{1} + m_{2})\cdot\bhh_{2} \\
        & \Rightarrow \bhh_{2} = \frac{m_{2} \cdot (n+m_{1}+m_{2})}{(1+\tau)\cdot(m_{1}+m_{2})} \,,
    \end{split}
\end{equation}

\noindent (throwing away the $\bhh_{2} = 0$ solution). By symmetry, we can send subscripted $1\rightarrow2$ and $2\rightarrow1$ to yield our $\bhh_{1}$ solution:

\begin{equation}
    \bhh_{1} = \frac{m_{1} \cdot (n+m_{1}+m_{2})}{(1+\tau)\cdot(m_{1}+m_{2})} \,.
\end{equation}

We now consider the more complex case where we have $\tau_{1} \neq \tau_{2}$ with $\tau_{1} > 0$, $\tau_{2} > 0$:

\begin{equation}
    \begin{split}
        0 & = n\cdot\bhh_{2} - (1+\tau_{2})\cdot(\bhh_{1}+\bhh_{2})\cdot\bhh_{2}+m_{2}\cdot(\bhh_{1}+\bhh_{2}) \\ & = (m_{2} - (1+\tau_{2})\cdot\bhh_{2})\cdot\bhh_{1} + ((n+m_{2}) - (1+\tau_{2})\cdot\bhh_{2})\cdot\bhh_{2} \\ & = (m_{2} - (1+\tau_{2})\cdot\bhh_{2})\cdot m_{1} + ((n+m_{2}) - (1+\tau_{2})\cdot\bhh_{2})\cdot(m_{2} + \Delta\tau_{12}\cdot\bhh_{2}) \\ & = m_{2}\cdot(m_{1}+m_{2}+n) + (\Delta\tau_{12}\cdot (n+m_{2}) - (1+\tau_{2})\cdot(m_{1}+m_{2}))\cdot\bhh_{2} - \Delta\tau_{12}\cdot(1+\tau_{2})\cdot\bhh_{2}^{2} \,,
    \end{split}
\end{equation}

\noindent where $\Delta\tau_{12} \equiv \tau_{1} - \tau_{2}$, which can only vary between $\tau_{1}$ and $-\tau_{2}$. We have also cancelled an overall factor of $\bhh_2$ on the third line, to remove the uninteresting solution $\bhh_2 = 0$. Our solution is:

\begin{equation}
    \bhh_{2} = \frac{-B_{2} \pm \sqrt{B_{2}^{2} - 4 A_{2}C_{2}}}{2A_{2}} \,,
\end{equation}

\noindent where:

\begin{equation}
    \begin{split}
        A_{2} & = -\Delta\tau_{12}\cdot(1+\tau_{2}) \,, \\
        B_{2} & = \Delta\tau_{12}\cdot (n+m_{2}) - (1+\tau_{2})\cdot(m_{1}+m_{2}) \,, \\
        C_{2} & = m_{2}\cdot(m_{1}+m_{2}+n) \,.
    \end{split}
\end{equation}

\noindent By symmetry, we have:

\begin{equation}
    \bhh_{1} = \frac{-B_{1} \pm \sqrt{B_{1}^{2} - 4 A_{1}C_{1}}}{2A_{1}} \,,
\end{equation}

\noindent where:

\begin{equation}
    \begin{split}
        A_{1} & = \Delta\tau_{12}\cdot(1+\tau_{1}) \,, \\
        B_{1} & = -\Delta\tau_{12}\cdot (n+m_{1}) - (1+\tau_{1})\cdot(m_{1}+m_{2}) \,, \\
        C_{1} & = m_{1}\cdot(m_{1}+m_{2}+n) \,.
    \end{split}
\end{equation}

The expressions above are only physically meaningful if $\bhh_{1} > 0$ and $\bhh_{2} > 0$ (you don't expect a \textit{negative} number of events). We suppose, for definiteness, $\tau_{1} > \tau_{2} \Rightarrow \Delta\tau_{12} > 0$ and \textit{require} $\bhh_{1} > 0$ and $\bhh_{2} > 0$ (i.e., we assume to have a physically meaningful solution). Then $A_2<0$ and $C_2>0$ so $-4A_{2}C_{2}>0$, which implies $B_{2}^2-4A_{2}C_{2}>0$ and $\sqrt{B_{2}^2-4A_{2}C_{2}}$ has a real root. Additionally, $2A_2<0$ and $B_2<\sqrt{B_{2}^2-4A_{2}C_{2}}$, so to always pick up a positive solution for $\bhh_{2}$, we choose the negative sign:

\begin{equation}
    \label{eqn:2cr_b2}
    \begin{split}
        \bhh_{2} & = \frac{-B_{2} - \sqrt{B_{2}^{2} - 4A_{2}C_{2}}}{2A_{2}} \\
        & = \frac{B_{2} + \sqrt{B_{2}^{2} + 4|A_{2}C_{2}|}}{2|A_{2}|} \,.
    \end{split}
\end{equation}

\noindent For $\Delta\tau_{12}>0$; this solution is real and positive. We turn to $\bhh_{1}$: $A_1>0$ and $C_1>0$ so $-4A_{1}C_{1}<0$ and $B_1 > \sqrt{B_1^2 - 4A_{1}C_{1}}$. Additionally, $-B_1>0$, so our solution is always positive and we may write it as:

\begin{equation}
    \bhh_{1} = \frac{|B_{1}| \pm \sqrt{B_{1}^{2} - 4|A_{1}C_{1}|}}{2|A_{1}|} \,.
\end{equation}

\noindent The sign choice is still ambiguous, so we return to Eq.~\ref{eqn:1cr_system_of_eqns} and subsitute in our expressions for each. One can show that the negative sign is required to solve our system of equations, and so our solution is:

\begin{equation}
    \label{eqn:2cr_b1}
    \bhh_{1} = \frac{|B_{1}| - \sqrt{B_{1}^{2} - 4|A_{1}C_{1}|}}{2|A_{1}|} \,.
\end{equation}

\noindent The above is always positive, but the condition for being real requires $B_{1}^{2} - 4|A_{1}C_{1}| > 0$. It can be shown that $B_{1}^{2} - 4A_{1}C_{1} = B_{2}^{2} - 4A_{2}C_{2}>0$ and so the real requirement is always met.

Our significance of discovery in the asymptotic limit is then Eq.~\ref{eqn:ncr_z0_diag} with Eqs.~\ref{eqn:2cr_b2} and \ref{eqn:2cr_b1} appropriately substituted in. Assuming Asimov data, we let $n = s + b_1 + b_2$, $m_1=\tau_1\cdot b_1$, and $m_2=\tau_2\cdot b_2$, where $s$ and $b_1$, $b_2$ are our theoretical signal and background yields in our SR, respectively.

\begin{figure}[htbp]
    \centering
    \includegraphics[width=0.9\textwidth]{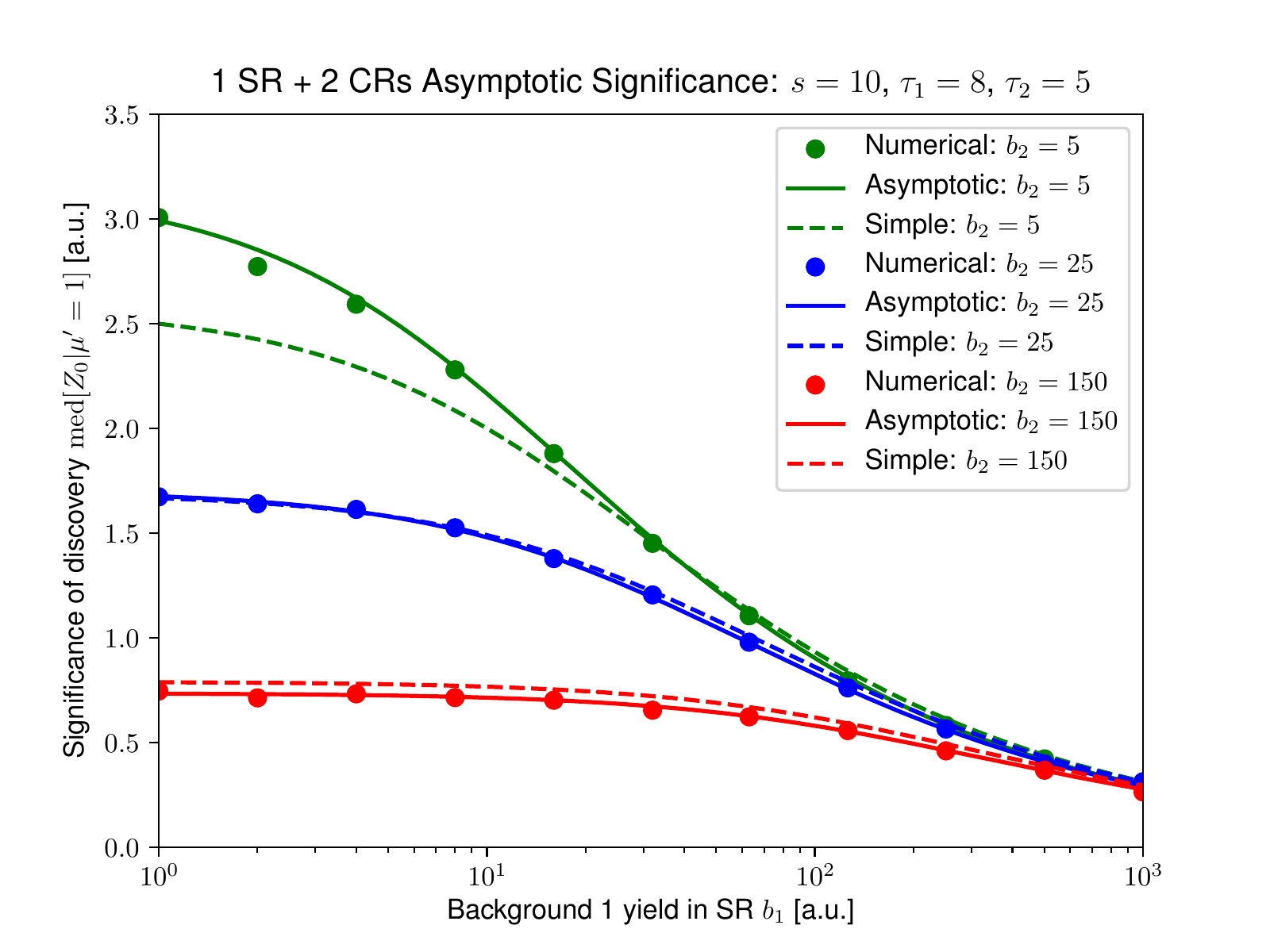}
    \caption{The median significance of discovery as a function of SR background 1 yield ($b_1$) and SR background 2 yield ($b_2$) for the 1 SR bin + 2 CR bins measurement described in Section~\ref{sec:1SRNCR_Neq2_diag_tau}. The SR signal yield $s$ is assumed to be 10. The transfer matrix for the background to the respective control regions is assumed to be diagonal with $\tau_1=8$ and $\tau_2=5$. ``Numerical'' refers to the results calculated using toy-based data (50,000 events for the estimation of $f(t_0|\mup=0)$ and 50,000 events for the estimation of $f(t_0|\mup=1)$, per point), ``Asymptotic'' refers to Eq.~\ref{eqn:ncr_z0_diag}, and ``Simple'' refers to $s/\sqrt{s+b_1+b_2}$.}
    \label{fig:1SR2CR}
\end{figure}

We have numerically calculated the median significance of discovery using the procedure described in Section~\ref{sec:simulation} and we have plotted Eq.~\ref{eqn:ncr_z0_diag} continuously alongside these numerical results: both the numerical and the asymptotic results are shown in Fig.~\ref{fig:1SR2CR}. Excellent agreement is observed, even down to low values of $s+b_1+b_2$. The ``naive'' approximation of the significance, $s/\sqrt{s+b_1+b_2}$, is also plotted. As expected, this naive approximation agrees well with the asymptotic and numerical results in the regime where $s/b \ll 1$ and diverges outside of that regime, as $s/\sqrt{s+b_1+b_2}$ is a Taylor expansion of the aymptotic result in the small $s/b$ limit~\cite{Cowan2011}. This is demonstrated most prominently by the green curve ($b_2 = 5$) at low values of $b_1$, where $s$, $b_1$, and $b_2$ are all $\mathcal{O}(1)$ and the $s/b \ll 1$ assumption fails.

\begin{figure}[htbp]
    \centering
    \begin{subfigure}{.49\textwidth}
        \centering
        \includegraphics[width=1.1\linewidth]{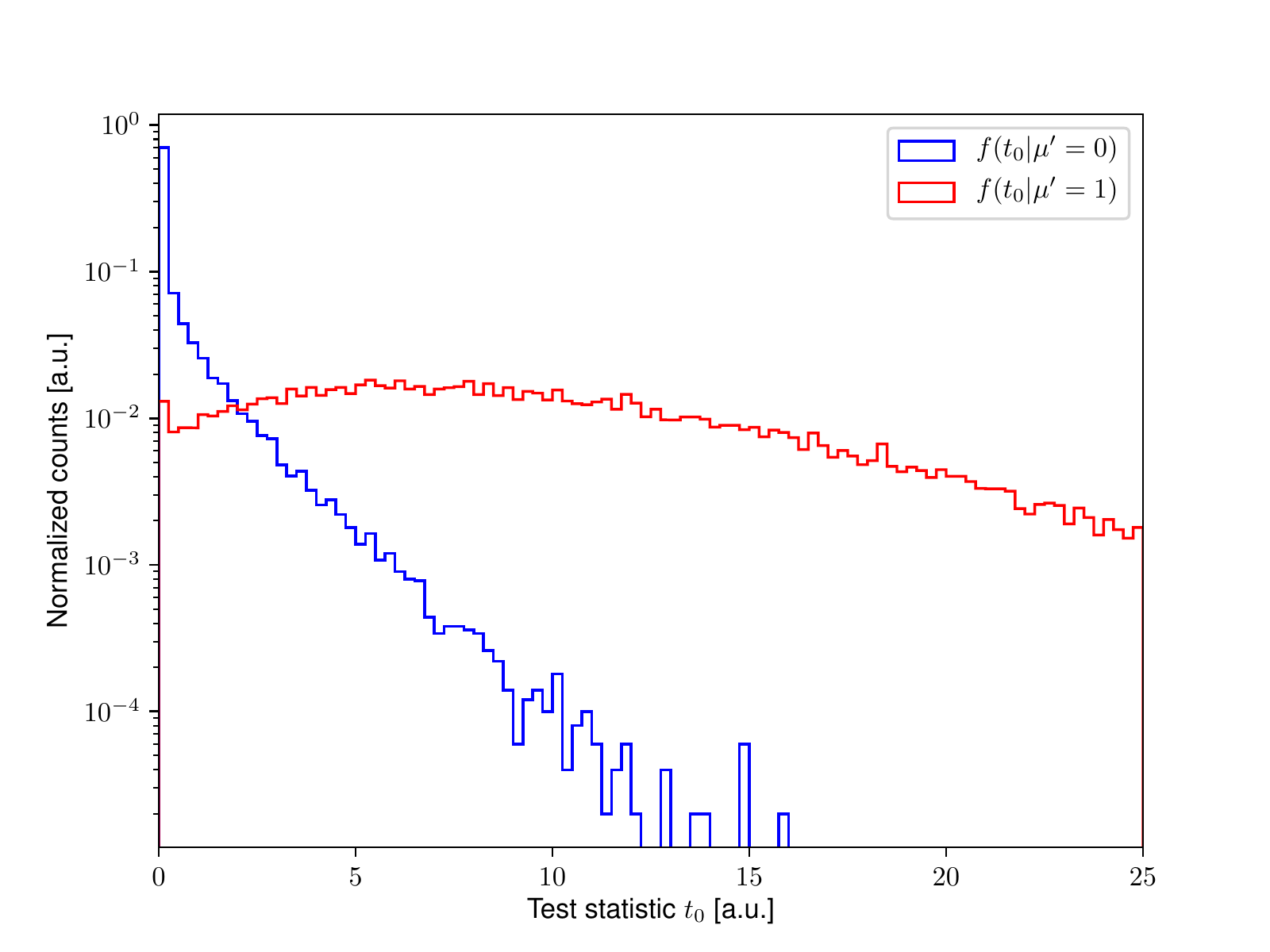}
        \caption{$b_1=1$.}
        \label{fig:2cr_diag_pdfs_1}
    \end{subfigure}
    \begin{subfigure}{.49\textwidth}
        \centering
        \includegraphics[width=1.1\linewidth]{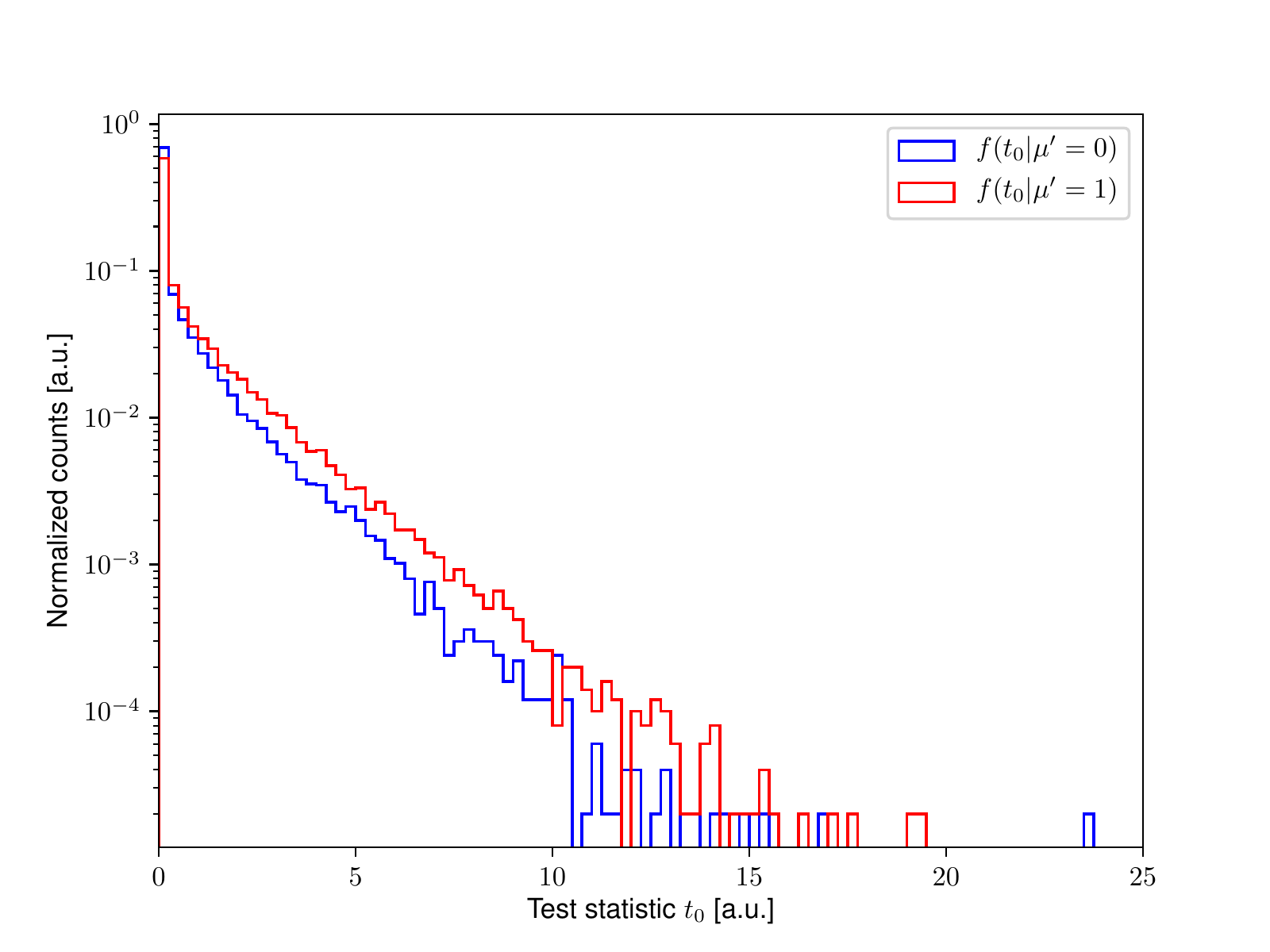}
        \caption{$b_1=1000$.}
        \label{fig:2cr_diag_pdfs_1000}
    \end{subfigure}
    \caption{PDFs of the test statistic of discovery $t_0$ for the green curve, $b_2 = 5$, in Fig.~\ref{fig:1SR2CR} for both the $b_1=1$ and $b_1=1000$ simulated data points. These distributions are used to calculate the corresponding values of $Z_0$ in Fig.~\ref{fig:1SR2CR}.}
    \label{fig:2cr_diag_pdfs}
\end{figure}

We also includes examples of the PDFs for our test statistic $t_0$ under the assumptions of no signal, $f(t_0|\mup=0)$, and in the presence of signal, $f(t_0|\mup=1)$, for the green curve, $b_2 = 5$, in Fig.~\ref{fig:1SR2CR} for both the $b_1 = 1$ and $b_1 = 1000$ simulated data points. As expected, $f(t_0|\mup=0)$ peaks at $t_0 = 0$ with a sharply falling tail. At higher values of $s/b$ as shown in Fig.~\ref{fig:2cr_diag_pdfs_1}, the median value of $f(t_0|\mup=1)$ is well offset from $t_0 =0 $, resulting in a smaller integrated $p$-value for the null hypothesis. At smaller values of $s/b$ as shown in Fig.~\ref{fig:2cr_diag_pdfs_1000}, the median value of $f(t_0|\mup=1)$ is approximately at $t_0 = 0$ and the distribution itself is not unlike $f(t_0|\mup=0)$, resulting in a larger integrated $p$-value. This behaviour is as expected.

\FloatBarrier

\subsection{$N$ Signal Regions + 1 Control Region}
\label{sec:NSR1CR}
We now consider the case where we have $N$ SRs (read as: $N$ signal bins) and 1 shared CR. Then for our $i=1,\ldots,N$ SRs, we have have transfer factors $\{\tau_i \,;\, i=1,\ldots,N\}$ where $\tau_i$ is the transfer factor carrying the background yield in our CR to SR $i$. We also assume the signal yields $s_1, s_2, \ldots, s_N$ among our $N$ SRs are correlated and tuned by a single POI, our signal strength $\mu$. Taking our theoretical background yield in our CR to be $b$ and our observed value to be $m$, we can write our likelihood as:

\begin{equation}
    L(\mu,b) = \prod_{i=1}^{N}\cbr{ \prob{n_i}{\mu s_i + \frac{b}{\tau_i}}} \cdot \prob{m}{b} \,,
\end{equation}

\noindent where we are dividing by the transfer factors, as in Section~\ref{sec:1SRNCR} we took $\tau$ to be the factor which multiplies yields in our SR to give yields in our CR. We can immediately write our log-likelihood as:

\begin{equation}
    \ln{L(\mu,b)} = \sum_{i=1}^{N}\br{n_i \cdot \ln\!\br{\mu s_i + \frac{b}{\tau_i}} - \mu s_i - \frac{b}{\tau_i}} + m\cdot\ln(b) - b \,,
\end{equation}

\noindent where we have discarded the constant $-(\ln(m!)+\sum_{i=1}^{N}\ln(n_i!))$. Going right ahead with finding our conditional and unconditional MLEs:

\begin{equation}
    \label{eqn:nsr_bhh}
    \begin{split}
        & \pd{\ln L(0,b)}{b}\bigg|_{b=\doublehatsubscript{b}} = \sum_{i=1}^{N}\br{\frac{n_i}{\bhh} - \frac{1}{\tau_i}} + \frac{m}{\bhh} - 1 = 0 \\ & \Leftrightarrow \frac{m + \sum_{i=1}^N n_i}{\bhh} = 1 + \sum_{i=1}^N \frac{1}{\tau_i} \\ & \Leftrightarrow \bhh = \frac{m + \sum_{i=1}^N n_i}{1 + \sum_{i=1}^N \frac{1}{\tau_i}} \,.
    \end{split}
\end{equation}

\noindent Also:

\begin{equation}
    \label{eqn:nsr_dLdmu}
    \pd{\ln L(\mu,b)}{\mu}\bigg|_{(\mu,b)=(\hat{\mu},\bh)} = \sum_{i=1}^{N}s_i\cdot\br{\frac{n_i}{\hat{\mu} s_i+\frac{\bh}{\tau_i}} - 1} = 0 \,,
\end{equation}

\noindent and:

\begin{equation}
    \label{eqn:nsr_dLdbh}
    \pd{\ln L(\mu,b)}{b}\bigg|_{(\mu,b)=(\hat{\mu},\bh)} = \sum_{i=1}^{N}\br{\frac{n_i}{\tau_i\hat{\mu}s_i+\bh} - \frac{1}{\tau_i}} + \frac{m}{\bh} - 1 = 0 \,.
\end{equation}

While these equations are difficult to solve in the general sense, we may propose the ansatz that $\hat{\mu}=1$ and $\bh=b$ when assuming Asimov data (i.e., $n_i = s_i + b/\tau_i\,\forall\,i=1,\ldots,N$ and $m=b$). While the solution is intuitive, it can be explicitly checked to solve Eq.~\ref{eqn:nsr_dLdmu}:

\begin{equation}
    \sum_{i=1}^{N}s_i\cdot\br{\frac{n_i}{\hat{\mu} s_i+\frac{\bh}{\tau_i}} - 1} = \sum_{i=1}^{N}s_i\cdot\br{\frac{s_i+\frac{b}{\tau_i}}{s_i+\frac{b}{\tau_i}} - 1} = \sum_{i=1}^{N}s_i\cdot\br{1 - 1} = 0 \,,
\end{equation}

\noindent and explicitly checked to solve Eq.~\ref{eqn:nsr_dLdbh}:

\begin{equation}
    \sum_{i=1}^{N}\br{\frac{n_i}{\tau_i\hat{\mu}s_i+\bh} - \frac{1}{\tau_i}} + \frac{m}{\bh} - 1 = \sum_{i=1}^{N}\br{\frac{s_i + \frac{b}{\tau_i}}{\tau_i s_i+b} - \frac{1}{\tau_i}} + \frac{b}{b} - 1 = \sum_{i=1}^{N}\br{\frac{1}{\tau_i} - \frac{1}{\tau_i}} + 1 - 1 = 0 \,.
\end{equation}

Using Eq.~\ref{eqn:Z0_simple} and the above solutions for $\hat{\mu}$ and $\bh$, we can write our significance of discovery in the asymptotic limit as:

\begin{equation}
    \label{eqn:nsr_z0}
    Z_0 = \sqrt{-2\cdot\br{ \sum_{i=1}^{N}\br{\br{s_i+\frac{b}{\tau_i}} \cdot \ln\!\br{\frac{\bhh}{\tau_i s_i + b}} + s_i + \frac{b-\bhh}{\tau_i}} + b\cdot\ln\!\br{\frac{\bhh}{b}} +(b - \bhh) }} \,,
\end{equation}

where $\bhh$ is given by Eq.~\ref{eqn:nsr_bhh} and Asimov data is assumed.

\begin{figure}[htbp]
    \centering
    \includegraphics[width=0.9\textwidth]{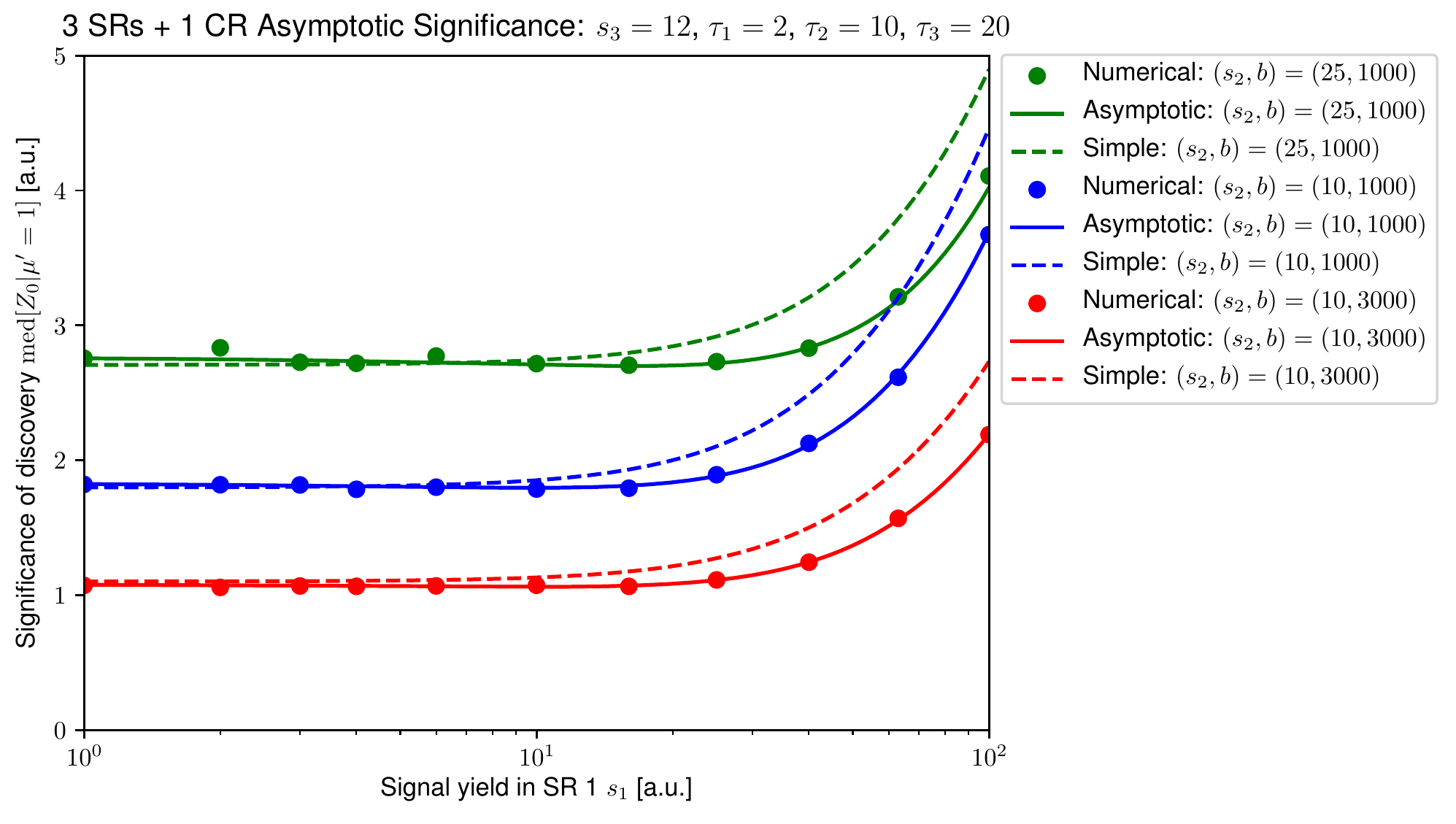}
    \caption{The median significance of discovery as a function of SR 1 signal yield ($s_1$), SR 2 signal yield ($s_2$), and CR background yield ($b$) for 3 SR bins + 1 CR bin measurement described in Section~\ref{sec:NSR1CR}. The SR 3 signal yield $s_3$ is assumed to be 12. The transfer factors for the background from the CR to SRs 1 ($\tau_1$), 2 ($\tau_2$), and 3 ($\tau_3$) are assumed to be 2, 10, and 20, respectively. ``Numerical'' refers to the results calculated using toy-based data (50,000 events for the estimation of $f(t_0|\mu^\prime=0)$ and 50,000 events for the estimation of $f(t_0|\mu^\prime=1)$, per point), ``Asymptotic'' refers to Eq.~\ref{eqn:nsr_z0}, and ``Simple'' refers to $\sqrt{\sum_{i=1}^N s_i^2/(s_i+b/\tau_i)}$. N.B.: the last data point for each curve, $s_1 = 100$, was simulated with 100,000 events for each PDF to ensure sufficient statistics for the $p$-value calculation.}
    \label{fig:3SR1CR}
\end{figure}

As in Section~\ref{sec:1SRNCR_Neq2_diag_tau}, we have numerically validated our results for the scenario where we have 3 SRs, $N=3$, and 1 CR, sampling Poisson PDFs in each of our 4 bins (with mean values of $s_1 + b/\tau_1$, $s_2 + b/\tau_2$, $s_3 + b/\tau_3$, and $b$) in order to generate our simulated yields. We plotted the asymptotic signficance of discovery, Eq.~\ref{eqn:nsr_z0}, continuously alongside these numerical results. This is shown in Fig.~\ref{fig:3SR1CR}. As before, we see excellent agreement between the numerical and asymptotic results over the range of theoretical yields and parameters studied. 

We have also plotted alongside our results the ``naive'' approximation of the significance of discovery where the bin-by-bin significances are summed in quadrature:

\begin{equation}
    Z_0 = \sqrt{\sum_{i=1}^N \br{\frac{s_i}{\sqrt{s_i+b/\tau_i}}}^2} = \sqrt{\sum_{i=1}^N \frac{s_i^2}{s_i+b/\tau_i}} \,,
\end{equation}

\noindent and indeed in the low $s/b$ regime (where we are in this regime if \textit{all} bins are in this regime), we see good agreement between all three methods. In the $s/b \sim 1$ regime (where we are in this regime if \textit{any} bin is in this regime), the naive approximation fails and no longer shows good agreement with the asymptotic and numerical methods, as expected.

\FloatBarrier

\subsection{$N$ Signal Regions + $M$ Control Regions}
\label{sec:NSRMCR}
We now consider the case where we have $N$ SRs (read as: $N$ signal bins) and $M$ CRs (one for each background process). We assume the signal yields $s_1, s_2, \ldots, s_N$ among our $N$ SRs are correlated and tuned by a single POI, our signal strength $\mu$. We also assume the definitions of the CRs are \textit{SR-independent} and \textit{orthogonal}. For $i=1,\ldots,N$ SRs, we have have transfer matrices $\{\boldsymbol{\tau}^i \,;\, i=1,\ldots,N\}$ where $\boldsymbol{\tau}^i$ is the transfer matrix carrying the yields in SR $i$ to our $M$ CRs (e.g., $\tau_{jk}^i$ carries the yield for background $k$ in SR $i$ to CR $j$). Our likelihood is:

\begin{equation}
    \label{eqn:nsr_mcr_L}
    L(\mu,\boldsymbol{B}) = \prod_{i=1}^{N}\cbr{ \prob{n_i}{\mu s_i + \sum_{j=1}^{M}b_{ji}}} \cdot \prod_{j=1}^{M} \cbr{\prob{m_j}{\sum_{j^{\prime}=1}^{M} \tau_{jj^{\prime}}^{i^\prime} b_{j^{\prime}i^{\prime}}}} \,,
\end{equation}

\noindent where $\boldsymbol{B}$ is our matrix of background yields in our SRs, where each row corresponds to specific background process and each column corresponds to a specific SR (e.g., $b_{ji} \equiv [\boldsymbol{B}]_{ji}$ corresponds to the yield for background $j$ in SR $i$). Additionally, $i^{\prime}$ can be any integer from 1 to $N$, but for consistent background yields in a given CR, regardless of which SR we are extrapolating from, we require $\sum_{j^{\prime}=1}^{M} \tau_{jj^{\prime}}^{i^\prime} b_{j^{\prime}i^{\prime}} = S_j \,\forall\, i^{\prime} = 1,\ldots,N$, where $S_j$ is the expected sum of weights in CR $j$ (i.e., a \textit{constant}). For definiteness, we take $i^\prime = 1$. Then taking the logarithm of Eq.~\ref{eqn:nsr_mcr_L}:

\begin{equation}
    \ln L(\mu,\boldsymbol{B}) = \sum_{i=1}^{N}\cbr{n_i\cdot\ln\!\br{\mu s_i + \sum_{j=1}^{M}b_{ji}} - s_i - \sum_{j=1}^{M}b_{ji} } + \sum_{j=1}^{M} \cbr{m_j\cdot\ln\!\br{\sum_{j^{\prime}=1}^{M} \tau_{jj^{\prime}}^{1} b_{j^{\prime}1}} - \sum_{j^{\prime}=1}^{M} \tau_{jj^{\prime}}^{1} b_{j^{\prime}1}} \,,
\end{equation}

\noindent where we have dropped the constant $-\sum_{i=1}^{N}\ln{n_i!} - \sum_{j=1}^{M}\ln{m_j!}$. Prior to taking any derivatives, we note:

\begin{equation}
    \pd{b_{ji}}{b_{\ell k}} = \delta_{j\ell} \cdot \pd{b_{ji}}{b_{jk}} = \delta_{j\ell} \cdot \frac{\tau^k_{1j}}{\tau^i_{1j}} \,,
\end{equation}

\noindent as two backgrounds from the same ``source'' (e.g., top quark pair production) maybe be linked via transfer factors, but never for backgrounds from different sources (e.g., top quark pair versus diboson production) will never be linked in this way. Additionally, we have used the fact that the background $j$ yield in SR $i$, $b_{ji}$, should give the same extrapolated background $j$ yield in CR 1 as the background $j$ yield in SR $k$, $b_{jk}$: $\tau^i_{1j}b_{ji} = \tau^k_{1j}b_{jk}$. N.B.: CR 1 was chosen for definiteness -- any of CRs $1,\ldots,M$ would work. Going ahead:

\begin{equation}
    \label{eqn:nsr_mcr_dLdbhh}
    \begin{split}
        & \pd{\ln L(\vec{0},\boldsymbol{B})}{b_{\ell k}}\bigg|_{\boldsymbol{B}=\doublehatsubscript{\boldsymbol{B}}} = \sum_{i=1}^{N}\frac{\tau^k_{1\ell}}{\tau^i_{1\ell}}\cdot\br{\frac{n_i}{\sum_{j=1}^{M}\bhh_{ji}} - 1} + \sum_{j=1}^{M} \frac{\tau^1_{j\ell}\tau^k_{1\ell}}{\tau^1_{1\ell}} \cdot \br{\frac{m_j}{\sum_{j^\prime=1}^{M}\tau^1_{jj^\prime}\bhh_{j^\prime 1}} - 1} = 0 \\
        & \Leftrightarrow \sum_{i=1}^{N}\frac{\tau^k_{1\ell}}{\tau^i_{1\ell}}\cdot\br{\frac{n_i}{\sum_{j=1}^{M}\bhh_{ji}} - 1} + \sum_{j=1}^{M} \frac{\tau^1_{j\ell}\tau^k_{1\ell}}{\tau^1_{1\ell}} \cdot \br{\frac{m_j}{[\boldsymbol{\tau}^1\cdot\doublehat{\boldsymbol{B}}]_{j1}} - 1} = 0 \,,
    \end{split}
\end{equation}

\noindent where we have compactified our denominator summation using matrix multiplication and also used:

\begin{equation}
\frac{\partial}{b_{\ell k}}\left[ \sum_{j=1}^{M}b_{ji} \right] = \sum_{j=1}^{M}\pd{b_{ji}}{b_{\ell k}} = \sum_{j=1}^{M}\delta_{j\ell}\cdot\frac{\tau^k_{1j}}{\tau^i_{1j}} = \frac{\tau^k_{1\ell}}{\tau^i_{1\ell}} \,,
\end{equation}

\noindent and:

\begin{equation}
\frac{\partial}{b_{\ell k}}\left[ \sum_{j^\prime=1}^{M}\tau^1_{jj^\prime}b_{j^\prime 1} \right] = \sum_{j^\prime=1}^{M}\tau^1_{jj^\prime} \cdot \pd{b_{j^\prime 1}}{b_{\ell k}} = \sum_{j^\prime=1}^{M}\tau^1_{jj^\prime} \cdot \delta_{j^\prime \ell} \cdot \frac{\tau^k_{1j^\prime}}{\tau^1_{1j^\prime}} = \frac{\tau^1_{j\ell}\tau^k_{1\ell}}{\tau^1_{1\ell}} \,.
\end{equation}

\noindent Given that $b_{ji}$ and $b_{jk}$ are \textit{not} independent, we can take our derivatives with respect to the background yields in only one of SRs $1,\ldots,N$ -- for definiteness, we choose SR 1 (i.e., $k=1$). So our best-fit values in the absence of signal are the solutions to the set of coupled equations:

\begin{equation}
    \label{eqn:nsr_mcr_bhh_eqns}
    \cbr{\sum_{i=1}^{N}\frac{\tau^1_{1\ell}}{\tau^i_{1\ell}}\cdot\br{\frac{n_i}{\sum_{j=1}^{M}\bhh_{ji}} - 1} + \sum_{j=1}^{M} \tau^1_{j\ell} \cdot \br{\frac{m_j}{[\boldsymbol{\tau}^1\cdot\doublehat{\boldsymbol{B}}]_{j1}} - 1} = 0 \,;\, \ell=1,{\ldots},M} \,.
\end{equation}

\noindent Also:

\begin{equation}
    \label{eqn:nsr_mcr_dLdmu}
    \pd{\ln L(\mu,\boldsymbol{B})}{\mu}\bigg|_{(\mu,\boldsymbol{B})=(\hat{\mu},\hat{\boldsymbol{B}})} = \sum_{i=1}^{N}s_i\cdot\br{\frac{n_i}{\hat{\mu} s_i+\sum_{j=1}^{M}\hat{b}_{ji}} - 1} = 0 \,,
\end{equation}

\noindent and:

\begin{equation}
    \label{eqn:nsr_mcr_dLdbh}
    \pd{\ln L(\mu,\boldsymbol{B})}{b_{\ell k}}\bigg|_{(\mu,\boldsymbol{B})=(\hat{\mu},\hat{\boldsymbol{B}})} = \sum_{i=1}^{N}\frac{\tau^k_{1\ell}}{\tau^i_{1\ell}}\cdot\br{\frac{n_i}{\hat{\mu}s_i + \sum_{j=1}^{M}\bhh_{ji}} - 1} + \sum_{j=1}^{M} \frac{\tau^1_{j\ell}\tau^k_{1\ell}}{\tau^1_{1\ell}} \cdot \br{\frac{m_j}{[\boldsymbol{\tau}^1\cdot\hat{\boldsymbol{B}}]_{j1}} - 1} = 0 \,.
\end{equation}

\noindent As in Section~\ref{sec:NSR1CR}, when we assume Asimov data, $n_i = s_i + \sum_{j=1}^{M}b_{ji} \,\forall\, i = 1,\ldots,N$ and $m_j = [\boldsymbol{\tau}^{1} \cdot \boldsymbol{B}]_{j1} \,\forall\, j = 1,\ldots,M$, we make the ansatz that our solutions are $\hat{\mu} = 1$ and $\hat{\boldsymbol{B}} = \boldsymbol{B}$ (N.B.: here, the elements of $\boldsymbol{B}$ are the theoretical background yields). This can be explicitly checked to solve Eq.~\ref{eqn:nsr_mcr_dLdmu}:

\begin{equation}
    \sum_{i=1}^{N}s_i\cdot\br{\frac{n_i}{\hat{\mu} s_i+\sum_{j=1}^{M}\hat{b}_{ji}} - 1} = \sum_{i=1}^{N}s_i\cdot\br{\frac{s_i + \sum_{j=1}^{M}b_{ji}}{s_i+\sum_{j=1}^{M}b_{ji}} - 1} = \sum_{i=1}^{N}s_i\cdot\br{1 - 1} = 0 \,,
\end{equation}

\noindent and explicitly checked to solve Eq.~\ref{eqn:nsr_mcr_dLdbh}:

\begin{equation}
    \begin{split}
        & \sum_{i=1}^{N}\frac{\tau^k_{1\ell}}{\tau^i_{1\ell}}\cdot\br{\frac{n_i}{\hat{\mu}s_i + \sum_{j=1}^{M}\bhh_{ji}} - 1} + \sum_{j=1}^{M} \frac{\tau^1_{j\ell}\tau^k_{1\ell}}{\tau^1_{1\ell}} \cdot \br{\frac{m_j}{[\boldsymbol{\tau}^1\cdot\hat{\boldsymbol{B}}]_{j1}} - 1} \\
        & = \sum_{i=1}^{N}\frac{\tau^k_{1\ell}}{\tau^i_{1\ell}}\cdot\br{\frac{s_i + \sum_{j=1}^{M}b_{ji}}{s_i + \sum_{j=1}^{M}\bhh_{ji}} - 1} + \sum_{j=1}^{M} \frac{\tau^1_{j\ell}\tau^k_{1\ell}}{\tau^1_{1\ell}} \cdot \br{\frac{[\boldsymbol{\tau}^1\cdot\boldsymbol{B}]_{j1}}{[\boldsymbol{\tau}^1\cdot\boldsymbol{B}]_{j1}} - 1} \\
        & = \sum_{i=1}^{N}\frac{\tau^k_{1\ell}}{\tau^i_{1\ell}}\cdot\br{1 - 1} +  \sum_{j=1}^{M} \frac{\tau^1_{j\ell}\tau^k_{1\ell}}{\tau^1_{1\ell}} \cdot \br{1 - 1} \\
        & = 0 \,.
    \end{split}
\end{equation}

Using Eq.~\ref{eqn:Z0_simple} and the above solutions for $\hat{\mu}$ and $\hat{\boldsymbol{B}}$, we can write our significance of discovery in the asymptotic limit as:

\begin{equation}
    \label{eqn:nsr_mcr_z0}
    Z_0 = \sqrt{-2\cdot\br{\sum_{i=1}^{N} \cbr{n_i\cdot\ln\!\br{\frac{\sum_{j=1}^{M}\bhh_{ji}}{n_i}} + n_i - \sum_{j=1}^{M}\bhh_{ji}} + \sum_{j=1}^{M} \cbr{ m_j\cdot\ln\!\br{\frac{[\boldsymbol{\tau}^1\cdot\doublehat{\boldsymbol{B}}]_{j1}}{[\boldsymbol{\tau}^1\cdot\boldsymbol{B}]_{j1}}} + [\boldsymbol{\tau}^1\cdot(\boldsymbol{B} - \doublehat{\boldsymbol{B}})]_{j1} }} } \,.
\end{equation}

\noindent where $\doublehat{\boldsymbol{B}}$ is given by Eq.~\ref{eqn:nsr_mcr_bhh_eqns} and Asimov data is assumed.

We have not provided numerical validation of the asymptotic results in this case, but we can show that the formulae reduce to the expected forms when there is only 1 SR, $N=1$. We let $s \equiv s_1$, $n \equiv n_1$, and $\boldsymbol{\tau} \equiv \boldsymbol{\tau}^1$. Additionally, we let $\boldsymbol{B} \rightarrow \bvec$ (i.e., $[\bvec]_j = [\boldsymbol{B}]_{j1} \,\forall\, j=1,\ldots,M$). In Eqs.~\ref{eqn:nsr_mcr_dLdbhh}~and~\ref{eqn:nsr_mcr_z0}, we have $[\boldsymbol{\tau}^1 \cdot \boldsymbol{B}]_{j1} \rightarrow [\boldsymbol{\tau} \cdot \bvec]_j = \sum_{\ell=1}^{M}\tau_{jj^\prime}b_{j^\prime}$ and similarly for $\doublehat{\boldsymbol{B}}$. Then Eq.~\ref{eqn:nsr_mcr_bhh_eqns} becomes:

\begin{equation}
    \frac{n}{\sum_{j=1}^{M}\bhh_j} - 1 + \sum_{j=1}^{M} \tau_{j\ell} \cdot \br{\frac{m_j}{\sum_{j^{\prime}=1}^{M} \tau_{jj^{\prime}} \bhh_{j^{\prime}}} - 1} = 0 \,,
\end{equation}

\noindent ($\forall\, \ell = 1,\ldots,M$) matching Eq.~\ref{eqn:ncr_bhh} (replacing $\ell \rightarrow k$ and $M \rightarrow N$), and Eq.~\ref{eqn:nsr_mcr_z0} becomes:

\begin{equation}
    Z_0 = \sqrt{-2\cdot\br{ n \cdot \ln\!\br{\frac{\sum_{j=1}^{M}\bhh_j}{n}} + n + \sum_{j=1}^{M}\br{-\bhh_{j} + m_{j} \cdot \ln\br{\frac{\sum_{j^\prime=1}^{M}\tau_{jj^\prime}\bhh_{j^\prime}}{\sum_{j^\prime=1}^{M}\tau_{jj^\prime}b_{j^\prime}}} + \sum_{j^\prime=1}^{M}\tau_{jj^\prime}\cdot(b_{j^\prime} - \bhh_{j^\prime})} }} \,.
\end{equation}

\noindent matching Eq.~\ref{eqn:ncr_z0} (replacing $M \rightarrow N$), as expected.

As additional validation, we can show that the formulae also reduce to the expected forms when there is only 1 CR, $M=1$. We let $b_i \equiv b_{1i} \,\forall\, i=1,\ldots,N$, $m \equiv m_1$, and $\tau_i \equiv [\boldsymbol{\tau}^i]_{11} \,\forall\, i=1,\ldots,N$. Additionally, we re-define the background yields in our SRs using the background yield in our 1 CR, $b$, letting $b_i \rightarrow b/\tau_i$ -- this is also implies $[\boldsymbol{\tau}^1 \cdot \boldsymbol{B}]_{j1} = \tau_1 b_1 \rightarrow b$. Then Eq.~\ref{eqn:nsr_mcr_bhh_eqns} becomes:

\begin{equation}
    \begin{split}
        & \sum_{i=1}^{N}\frac{\tau_1}{\tau_i}\cdot\br{\frac{n_i}{\bhh/\tau_i} - 1} + \tau_1 \cdot \br{\frac{m}{\bhh} - 1} = 0 \\
        & \Leftrightarrow m + \sum_{i=1}^{N}n_i = \bhh\cdot\br{1 + \sum_{i=1}^{N}\frac{1}{\tau_i}} \\
        & \Leftrightarrow \bhh = \frac{m + \sum_{i=1}^{N}n_i}{1 + \sum_{i=1}^{N}\frac{1}{\tau_i}} \,,
    \end{split}
\end{equation}

\noindent matching Eq.~\ref{eqn:nsr_bhh}, and Eq.~\ref{eqn:nsr_mcr_z0} becomes:

\begin{equation}
    Z_0 = \sqrt{-2\cdot\br{ \sum_{i=1}^{N} \br{n_i \cdot \ln\!\br{\frac{\bhh}{\tau_i n_i}} + n_i - \frac{\bhh}{\tau_i}} + m \cdot \ln\!\br{\frac{\bhh}{b}} + (b - \bhh)}} \,.
\end{equation}

\noindent matching Eq.~\ref{eqn:nsr_z0} after replacing $n_i = s_i + b/\tau_i \,\forall\, i=1,\ldots,N$ and $m=b$ (i.e., their Asimov values).

\FloatBarrier

\subsection{1 Signal Region + $M$ Gaussian Background Constraints}
\label{sec:gauss}
\subsubsection{General Case}
\label{sec:gauss_general}

We consider the case where we have 1 SR (read as: 1 bin) with a signal process yield $s$ and $N$ background processes with $M$ Gaussian constraints on those backgrounds (read as: $M$ NPs). We assume NP $j$, $\theta_j$, is described by a Gaussian constraint with a nominal value of 0, a mean of $\theta_j$, and a variance of 1 as well as that the NPs are related via a \textit{correlation} matrix $\boldsymbol{\Sigma}$. Additionally, we assume background process $i$ is affected by the NPs via:

\begin{equation}
    b_i \rightarrow \prod_{j=1}^M R_{ij}(\theta_j)\cdot b_i \,,
\end{equation}

\noindent where $R_{ij}(\theta_j)$ is the response function of background $i$ to NP $j$, as described in Ref.~\cite{Conway2011}. We'll condense our notation in the following equations by letting $R_i(\vec{\theta}) \equiv \prod_{j=1}^M R_{ij}(\theta_j)$. In our single SR bin, we assume the behaviour of a response function is governed by:

\begin{equation}
    \label{eqn:response}
    \begin{split}
        R_{ij}(+1) & = 1 + \sigma_{ij}^+ \,, \\
        R_{ij}(0)  & = 1  \,, \\
        R_{ij}(-1) & = 1 - \sigma_{ij}^- \,,
    \end{split}
\end{equation}

\noindent where $\sigma_{ij}^+$ and $\sigma_{ij}^-$ are constants dependent on the NP considered (i.e., NP $j$) and affecting the overall normalization of background $i$ (i.e., are \textit{relative} ``uncertainties''). The functional form of $R_{ij}(\theta_j)$ for $\theta_j \in [-1, +1]$ is left free. Note: $R_{ij}(\theta_j)=1$ if background~$i$ is \textit{not} affected by NP~$j$.

Our likelihood may be written as:

\begin{equation}
    L(s,\vec{\theta}) = \prob{n}{s + \sum_{i=1}^{N}R_i(\vec{\theta})\cdot b_{i}} \cdot G\br{\vec{0} \,\middle|\, \vec{\theta}, \boldsymbol{\Sigma}} \,,
\end{equation}

\noindent where we have moved the functional dependence from $\vec{b}$ to $\vec{\theta}$ (as the value of $\vec{\theta}$ tunes the background yields) and where the diagonal elements of $\boldsymbol{\Sigma}$ are assumed to be 1. If the NPs were fully decoupled from one another, $\boldsymbol{\Sigma}$ would be a diagonal matrix and our $M$-dimensional Gaussian function may be written as the product of $M$ 1-dimensional Gaussian constraints. Substituting in explicit expressions for our Poisson PDF and Gaussian constraint:

\begin{equation}
    L(s,\vec{\theta}) = \pois{n}{s + \sum_{i=1}^{N}R_i(\vec{\theta})\cdot b_{i}} \cdot \frac{\exp\br{-\frac{1}{2}\cdot\vec{\theta}^\top\boldsymbol{\Sigma}^{-1}\vec{\theta}}}{\sqrt{(2\pi)^M |\boldsymbol{\Sigma}|}} \,,
\end{equation}

\noindent or taking the logarithm:

\begin{equation}
    \ln L(s,\tvec) = n\cdot\ln\!\br{s + \sum_{i=1}^{N}R_i(\tvec)\cdot b_{i}} - s - \sum_{i=1}^{N}R_i(\tvec)\cdot b_{i} - \frac{1}{2}\cdot\tvec^\top\boldsymbol{\Sigma}^{-1}\tvec \,,
\end{equation}

\noindent where we dropped the constant $-(\ln(n!) + \ln(\sqrt{(2\pi)^M |\boldsymbol{\Sigma}|}))$. Note that we can also write the matrix multiplication in the last term using sums:

\begin{equation}
    \tvec^\top\boldsymbol{\Sigma}^{-1}\tvec = \sum_{i=1}^M\sum_{j=1}^M [\boldsymbol{\Sigma}^{-1}]_{ij}\theta_i\theta_j = \sum_{i=1}^M\sum_{j=i}^M (2 - \delta_{ij})[\boldsymbol{\Sigma}^{-1}]_{ij}\theta_i\theta_j \,,
\end{equation}

\noindent where $[\boldsymbol{\Sigma}^{-1}]_{ij}$ refers to the element in the $i$-th row and $j$-th column of $\boldsymbol{\Sigma}^{-1}$ and $\delta_{ij}$ is the Kronecker delta function. Here, we have used the fact that if $\boldsymbol{\Sigma}$ is symmetric then its inverse is also symmetric: $\boldsymbol{\Sigma}^{-1} = (\boldsymbol{\Sigma}^{-1})^\top$. From the above, it is also apparent that:

\begin{equation}
    \pd{(\tvec^\top\boldsymbol{\Sigma}^{-1}\tvec)}{\theta_k} = 2\cdot\sum_{j=1}^M [\boldsymbol{\Sigma}^{-1}]_{kj}\theta_j \,.
\end{equation}

Continuing ahead, we consider our best fit values in the absence of signal:

\begin{equation}
    \pd{\ln L(0,\tvec)}{\theta_k}\bigg|_{\tvec=\thhvec} = \br{\frac{n}{\sum_{i=1}^{N}R_i(\thhvec)\cdot b_{i}} - 1}\cdot \sum_{i=1}^N \cbr{\br{\pd{R_{ik}(\theta_k)}{\theta_k}\bigg|_{\theta_k=\doublehatsubscript{\theta}_k}} \cdot \frac{R_i(\thhvec)\cdot b_i}{R_{ik}(\thh_k)}} - \sum_{j=1}^M [\boldsymbol{\Sigma}^{-1}]_{kj}\thh_j = 0 \,,
\end{equation}

\noindent (the above is not unlike Eq.~8 of Ref.~\cite{Conway2011} for decorrelated constraints) and so our system of equations solving for $\thhvec$ are:

\begin{equation}
    \label{eqn:gauss_thetahh}
    \cbr{\br{\frac{n}{\sum_{i=1}^{N}R_i(\thhvec)\cdot b_{i}} - 1}\cdot \sum_{i=1}^N \cbr{\br{\pd{R_{ik}(\theta_k)}{\theta_k}\bigg|_{\theta_k=\doublehatsubscript{\theta}_k}} \cdot \frac{R_i(\thhvec)\cdot b_i}{R_{ik}(\thh_k)}} - \sum_{j=1}^M [\boldsymbol{\Sigma}^{-1}]_{kj}\thh_j = 0 \,;\, k = 1,\ldots,M} \,.
\end{equation}

We now consider the best fit values in the context of our tested hypothesis:

\begin{equation}
    \label{eqn:gauss_sh}
    \pd{\ln L(s,\tvec)}{s}\bigg|_{(s,\tvec)=(\sh,\thvec)} = \frac{n}{\sh+\sum_{i=1}^{N}R_i(\thvec)\cdot b_i} - 1 = 0 \Leftrightarrow \sh = n - \sum_{i=1}^{N}R_i(\thvec)\cdot b_i \,,
\end{equation}

\noindent and:

\begin{equation}
    \begin{split}
        \pd{\ln L(s,\tvec)}{\theta_k}\bigg|_{(s,\tvec)=(\sh,\thvec)} & = \br{\frac{n}{\sh + \sum_{i=1}^{N}R_i(\thvec)\cdot b_i} - 1}\cdot \sum_{i=1}^N \cbr{\br{\pd{R_{ik}(\theta_k)}{\theta_k}\bigg|_{\theta_k=\that_k}} \cdot \frac{R_i(\thvec)\cdot b_i}{R_{ik}(\that_k)}} - \sum_{j=1}^M [\boldsymbol{\Sigma}^{-1}]_{kj}\that_j = 0 \\ & \Rightarrow \sum_{j=1}^M [\boldsymbol{\Sigma}^{-1}]_{kj}\that_j = 0 \,,
    \end{split}
\end{equation}

\noindent where we substituted in Eq.~\ref{eqn:gauss_sh}. The above can be simultaneously satisfied for all $k$ by writing $\boldsymbol{\Sigma}^{-1}\thvec = \vec{0} \Rightarrow \thvec = \vec{0}$. Accordingly, $\sh = n - \sum_{i=1}^N R_i(\vec{0})\cdot b_i = n - \sum_{i=1}^N b_i$. Our significance of discovery in the asymptotic limit is thus:

\begin{equation}
    \label{eqn:gauss_z0}
    \begin{split}
        Z_0 & = \sqrt{-2\cdot\ln\!\br{\frac{L(0,\thhvec)}{L(\sh,\thvec)}}} \\
        & = \sqrt{ -2\cdot\br{n\cdot\ln\!\br{\frac{\sum_{i=1}^{N}R_i(\thhvec)\cdot b_{i}}{n}} + n - \sum_{i=1}^{N}R_i(\thhvec)\cdot b_{i} - \frac{1}{2}\cdot\thhvec^\top\boldsymbol{\Sigma}^{-1}\thhvec} } \,.
    \end{split}
\end{equation}

\noindent Assuming Asimov data, we let $n = s + \sumb$ in Eq.~\ref{eqn:gauss_z0} and appropriately substitute the solutions from Eq.~\ref{eqn:gauss_thetahh}.

\subsubsection{Assuming $N$ Backgrounds and $M=N$ Decorrelated Constraints, 1-Per-Background}
\label{sec:gauss_decorr}

We consider $N$ backgrounds with $M=N$ Gaussian contraints, one per background. We also assume the constraints are decorrelated (i.e., $\boldsymbol{\Sigma}=\boldsymbol{I}\Rightarrow\boldsymbol{\Sigma}^{-1}=\boldsymbol{I}$ where $\boldsymbol{I}$ is the identity matrix). Then $R_i(\vec{\theta}) \rightarrow R_i(\theta_i)$ and $R_{ij}(\theta_j) \rightarrow R_i(\theta_i)$ (i.e., the total response function for background $i$ is only a function of $\theta_i$, the NP governing the constraint which affects only background $i$), implying:

\begin{equation}
    \label{eqn:gauss_decorr_dRdth}
    \pd{R_{ik}(\theta_k)}{\theta_k} = \pd{R_{i}(\theta_i)}{\theta_k} = \delta_{ik}\cdot\pd{R_{k}(\theta_k)}{\theta_k} \,.
\end{equation}

\noindent Applying the Eq.~\ref{eqn:gauss_decorr_dRdth} to Eq.~\ref{eqn:gauss_thetahh} yields:

\begin{equation}
    \label{eqn:gauss_decorr_thetahh}
    \cbr{\br{\frac{n}{\sum_{i=1}^{N}R_i(\thh_i)\cdot b_{i}} - 1}\cdot \pd{R_{k}(\theta_k)}{\theta_k}\bigg|_{\theta_k=\doublehatsubscript{\theta}_k} \cdot b_k - \thh_k = 0 \,;\, k = 1,\ldots,N} \,.
\end{equation}

For an additional level of simplicity, we assume the response functions are linear in $\theta_k$ and that $\sigma_k^+ = \sigma_k^- \equiv \sigma_k$: $R_k(\theta_k) = \sigma_k\cdot\theta_k + 1$, where $\sigma_k$ can be interpreted as the \textit{relative} ``uncertainty'' on the normalization of background $k$. Consider the redefinitions: $R_k(\theta_k)\cdot b_k \rightarrow b_k(\theta_k)$ and $\sigma_k\cdot b_k \rightarrow \sigma_k$ (i.e., $\sigma_k$ is now an \textit{absolute} ``uncertainty''). Then we let $\bhh_k \equiv b_k(\thh_k)$ and:

\begin{equation}
    \label{eqn:gauss_decorr_dRdthetahh}
    \pd{R_k(\theta_k)}{\theta_k}\bigg|_{\theta_k=\doublehatsubscript{\theta}_k} \cdot b_k = \pd{(R_k(\theta_k)\cdot b_k)}{\theta_k}\bigg|_{\theta_k=\doublehatsubscript{\theta}_k} = \pd{b_k(\theta_k)}{\theta_k}\bigg|_{\theta_k=\doublehatsubscript{\theta}_k} = \sigma_k \,.
\end{equation}

\noindent We identify $\thh_k = (\bhh_k-b_k)/\sigma_k$ (i.e., $\thh_k$ controls the tuning of the best-fit $\bhh_k$ away from the nominal value $b_k$, consistent with our earlier definition of the response function), so inserting this and Eq.~\ref{eqn:gauss_decorr_dRdthetahh} into Eq.~\ref{eqn:gauss_decorr_thetahh}\footnote{This system of equations is (attempted to be) solved in general sense in Appendix~\ref{sec:appA}}:

\begin{equation}
    \label{eqn:gauss_decorr_bhh}
    \cbr{\br{\frac{n}{\sum_{i=1}^{N}\bhh_{i}} - 1}\cdot\sigma_k - \frac{(\bhh_k - b_k)}{\sigma_k} = 0 \,;\, k = 1,\ldots,N} \,,
\end{equation}

\noindent which can be solved to yield the best-fit values of the backgrounds in the absence of signal. Finally, with our assumptions and redefinitions, Eq.~\ref{eqn:gauss_z0} becomes:

\begin{equation}
    \label{eqn:gauss_decorr_z0}
    Z_0 = \sqrt{ -2\cdot\br{n\cdot\ln\!\br{\frac{\sum_{i=1}^{N}\bhh_{i}}{n}} + n - \sum_{i=1}^{N}\br{\bhh_{i} + \frac{(b_i-\bhh_i)^2}{2\sigma_i^2}}} } \,,
\end{equation}

\noindent which is our significance of discovery in the asymptotic limit. Assuming Asimov data, we would let $n = s + \sum_{i=1}^{N}b_i$.

\begin{figure}[htbp]
    \centering
    \includegraphics[width=0.9\textwidth]{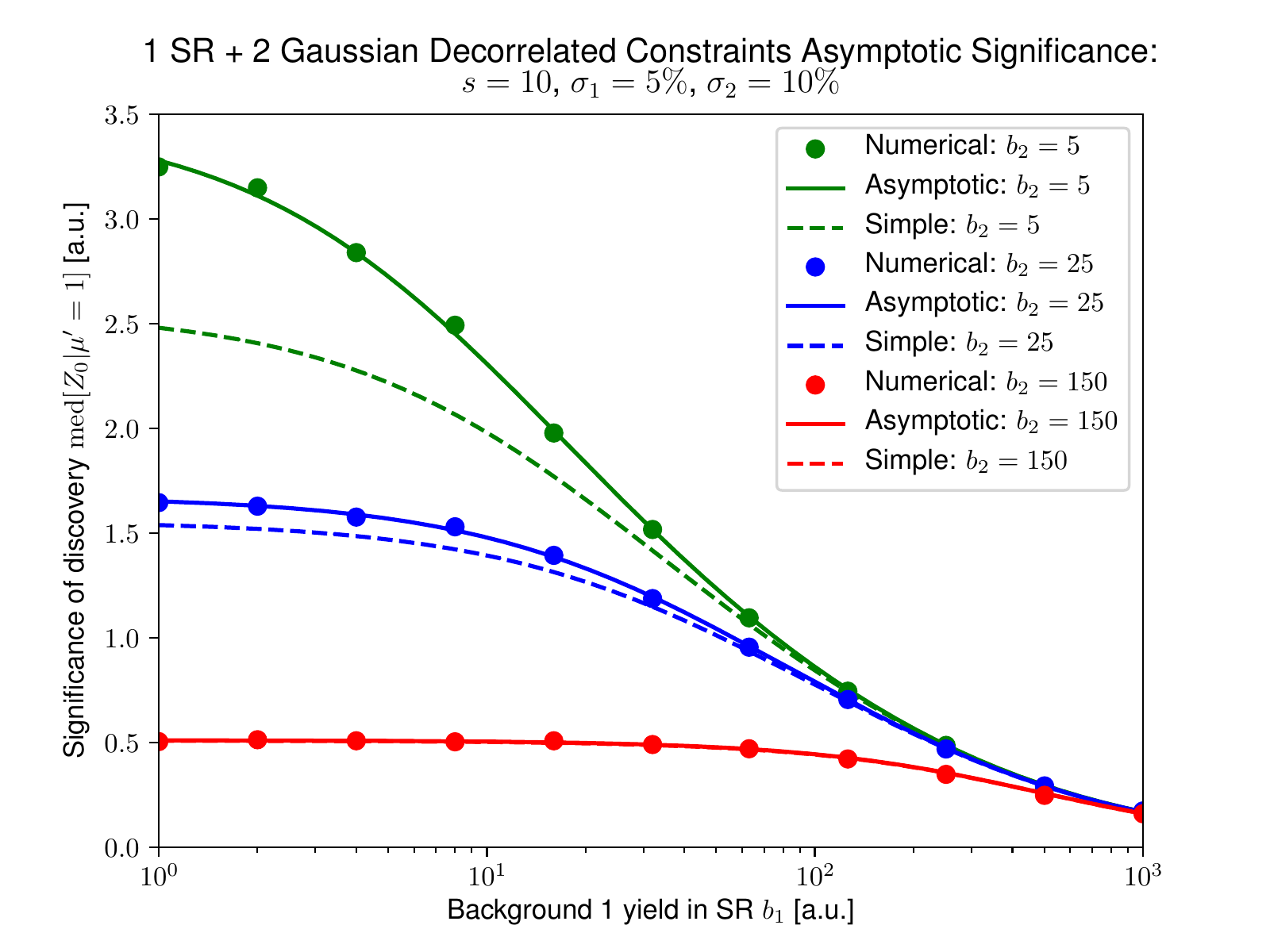}
    \caption{The median significance of discovery as a function of SR background 1 yield ($b_1$) and SR background 2 yield ($b_2$) for 1 SR bin + 2 decorrelated Gaussian constraints measurement described in Section~\ref{sec:gauss_decorr}. The SR signal yield is assumed to be 10. The relative uncertainties on backgrounds 1 and 2 are assumed to be $\sigma_1=5\%$ and $\sigma_2=10\%$, respectively. ``Numerical'' refers to the results calculated using toy-based data (50,000 events for the estimation of $f(t_0|\mu^\prime=0)$ and 50,000 events for the estimation of $f(t_0|\mu^\prime=1)$, per point), ``Asymptotic'' refers to Eq.~\ref{eqn:gauss_decorr_z0}, and ``Simple'' refers to $s/\sqrt{s+b_1+b_2+\sigma_1^2+\sigma_2^2}$.}
    \label{fig:1SR2GaussConst}
\end{figure}

\begin{figure}[htbp]
    \centering
    \hspace{0.2\textwidth}\includegraphics[width=0.75\linewidth]{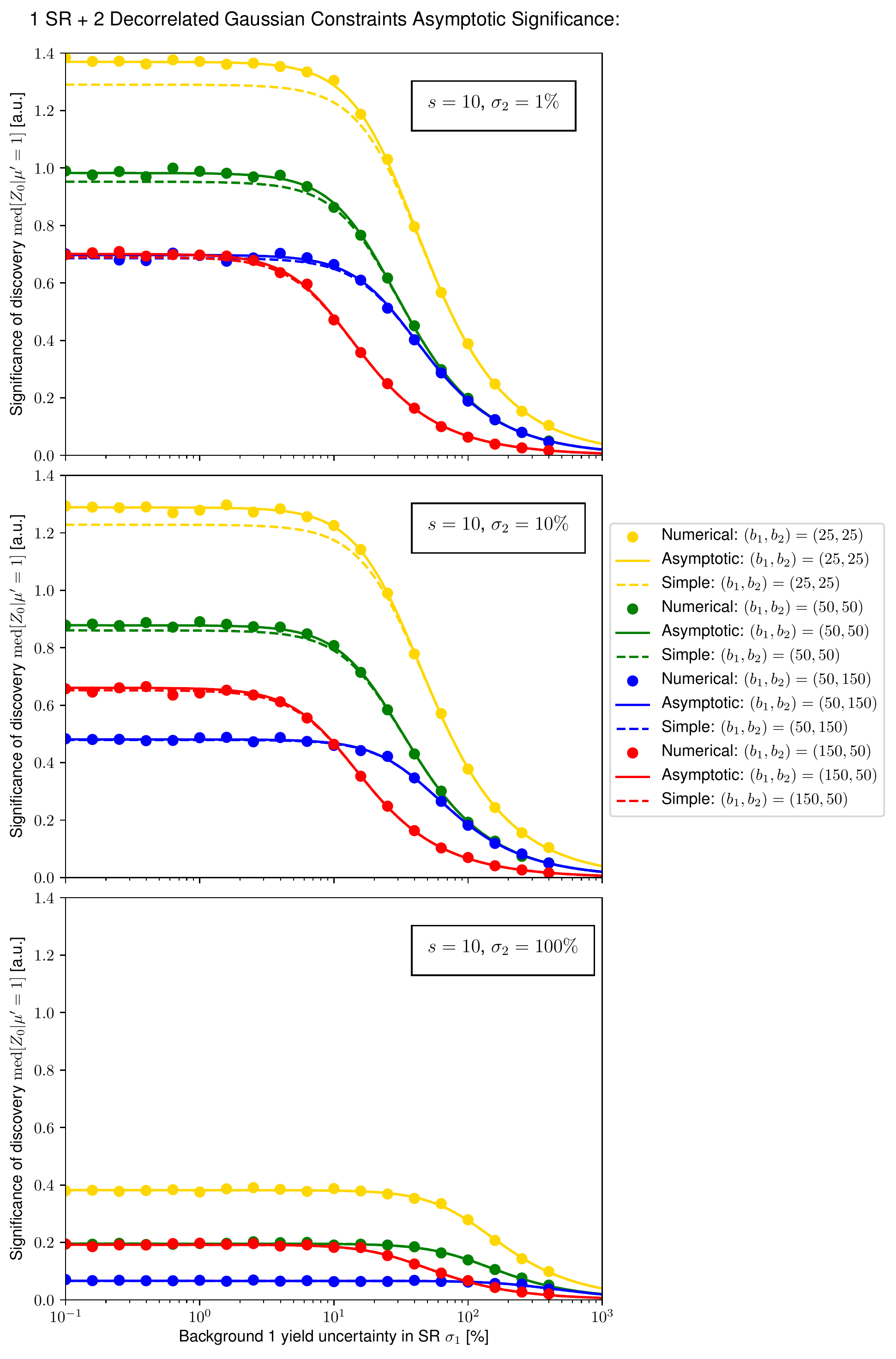}
    \caption{The median significance of discovery as a function of the relative uncertainty on the SR background 1 yield ($\sigma_1$), the SR background 1 yield ($b_1$), and the SR background 2 yield ($b_2$) for 1 SR bin + 2 decorrelated Gaussian constraints measurement described in Section~\ref{sec:gauss_decorr}. The SR signal yield is assumed to be 10. Different values for the relative uncertainty on the background 2 yield ($\sigma_2$) are also tested: (top) 1\%, (middle) 10\%, and (bottom) 100\%. ``Numerical'' refers to the results calculated using toy-based data (50,000 events for the estimation of $f(t_0|\mu^\prime=0)$ and 50,000 events for the estimation of $f(t_0|\mu^\prime=1)$, per point), ``Asymptotic'' refers to Eq.~\ref{eqn:gauss_decorr_z0}, and ``Simple'' refers to $s/\sqrt{s+b_1+b_2+\sigma_1^2+\sigma_2^2}$.}
    \label{fig:1SR2GaussConst_err}
\end{figure}

We have also numerically validated our results for the scenario where we have 2 backgrounds each with 1 Gaussian constraint -- we also assume the constraints are decorrelated. To simulate our yields, we sample a Poisson PDF with mean $\mu = s + b_1 + b_2$ in our SR. Additionally, to avoid biasing ourselves, we must sample the nominal value of each of the NPs from a Gaussian PDF centered on $b_i$ with width $\sigma_i$ for $i=1$ and then $i=2$. The sampled values then become the ``true'' nominal values of the Gaussian constraints used in our maximum likelihood fit (i.e., the constraints look like $G(\tilde{b}_i\,|\,b_i,\sigma_i)$ with $\tilde{b}_i$ sampled from $G(x\,|\,b_i,\sigma_i)$). We have plotted the asymptotic signficance of discovery, Eq.~\ref{eqn:gauss_decorr_z0}, continuously alongside these numerical results -- this is shown in Figs.~\ref{fig:1SR2GaussConst} and \ref{fig:1SR2GaussConst_err}. As before, we see excellent agreement between the numerical and asymptotic results over the range of theoretical yields and parameters studied, including $s/b$ from $\mathcal{O}(0.01)$ to $\mathcal{O}(1)$ and uncertainties on the background yields from $\mathcal{O}(0.1\%)$ to $\mathcal{O}(100\%)$.

Alongside our results, we have also plotted the ``naive'' approximation of the significance of discovery:

\begin{equation}
    Z_0 = \frac{s}{\sqrt{s + b_1 + b_2 + \sigma_1^2 + \sigma_2^2}} \,,
\end{equation}

\noindent (where $\sigma_1$ and $\sigma_2$ are given as the absolute uncertainties on backgrounds $1$ and $2$, respectively) and indeed in the low $s/b$ (i.e., high $b_1+b_2$) regime we see good agreement between all three methods. In the $s/b \sim 1$ regime (i.e., low $b_1+b_2$), the naive approximation fails and no longer shows good agreement with the asymptotic and numerical methods, as expected. We also note that in the high uncertainty regime, $\sigma \sim 100\%$, the naive, asymptotic, and numerical methods similarly agree well with one another, exemplified by the bottom plot in Fig.~\ref{fig:1SR2GaussConst_err}.

\subsubsection{Assuming $N=1$ Backgrounds and $M=1$ Constraints}
\label{sec:gauss_decorr_NeqMeq1}

As a special case of the previous section, we consider only $N=1$ backgrounds with $M=N=1$ Gaussian contraints on this background (with 1 background and 1 constraint, we are automatically in the regime of ``decorrelated'' constraints). Letting $b \equiv b_1$ and $\sigma \equiv \sigma_1$, Eq.~\ref{eqn:gauss_decorr_bhh} becomes:

\begin{equation}
    \begin{split}
        & \br{\frac{n}{\bhh} - 1}\cdot\sigma - \frac{(\bhh-b)}{\sigma} = 0 \\ & \Leftrightarrow \br{n-\bhh}\cdot\sigma^2 -(\bhh-b)\cdot\bhh = 0 \\ & \Leftrightarrow -\bhh^2 + (b-\sigma^2)\cdot\bhh + n\sigma^2 = 0 \\ & \Rightarrow \bhh = \frac{-(b-\sigma^2) \pm \sqrt{(b-\sigma^2)^2 + 4 n\sigma^2}}{-2} \,,
    \end{split}
\end{equation}

\noindent but as $(b-\sigma^2)^2 + 4 n\sigma^2 > (b-\sigma^2)^2$ (all variables in this expression are positive), we choose the minus sign to give an overall positive (i.e., physical) solution for $\bhh$:

\begin{equation}
    \label{eqn:gauss_simple_bhh}
    \bhh = \frac{(b-\sigma^2) + \sqrt{(b-\sigma^2)^2 + 4 n\sigma^2}}{2} \,.
\end{equation}

Using Eq.~\ref{eqn:gauss_decorr_z0}, our significance of discovery is:

\begin{equation}
    \label{eqn:gauss_simple_z0}
    Z_0 = \sqrt{ -2\cdot\br{n\cdot\ln\!\br{\frac{\bhh}{n}} + n - \bhh - \frac{(b-\bhh)^2}{2\sigma^2}} } \,,
\end{equation}

\noindent with Eq.~\ref{eqn:gauss_simple_bhh} appropriately substituted in, matching what is shown in Eq.~26 of Ref.~\cite{Buttinger2019}.

\subsubsection{Correlated Nusiance Parameters}
\label{sec:gauss_corr}

One feature which is not accounted for by the formulae available in the literature are correlations between the NPs -- these correlations are built into Eqs.~\ref{eqn:gauss_thetahh} and \ref{eqn:gauss_z0}. While the assumption of decorrelated NPs (i.e., $\boldsymbol{\Sigma} = \boldsymbol{I}$) is applicable to most practical use cases, it is interesting to examine how the correlations affect the estimated sensitivity. The expectation is that introducing correlations will \textit{decrease} the sensitivity relative to the decorrelated regime.

\begin{figure}[htbp]
    \centering
    \includegraphics[width=0.9\linewidth]{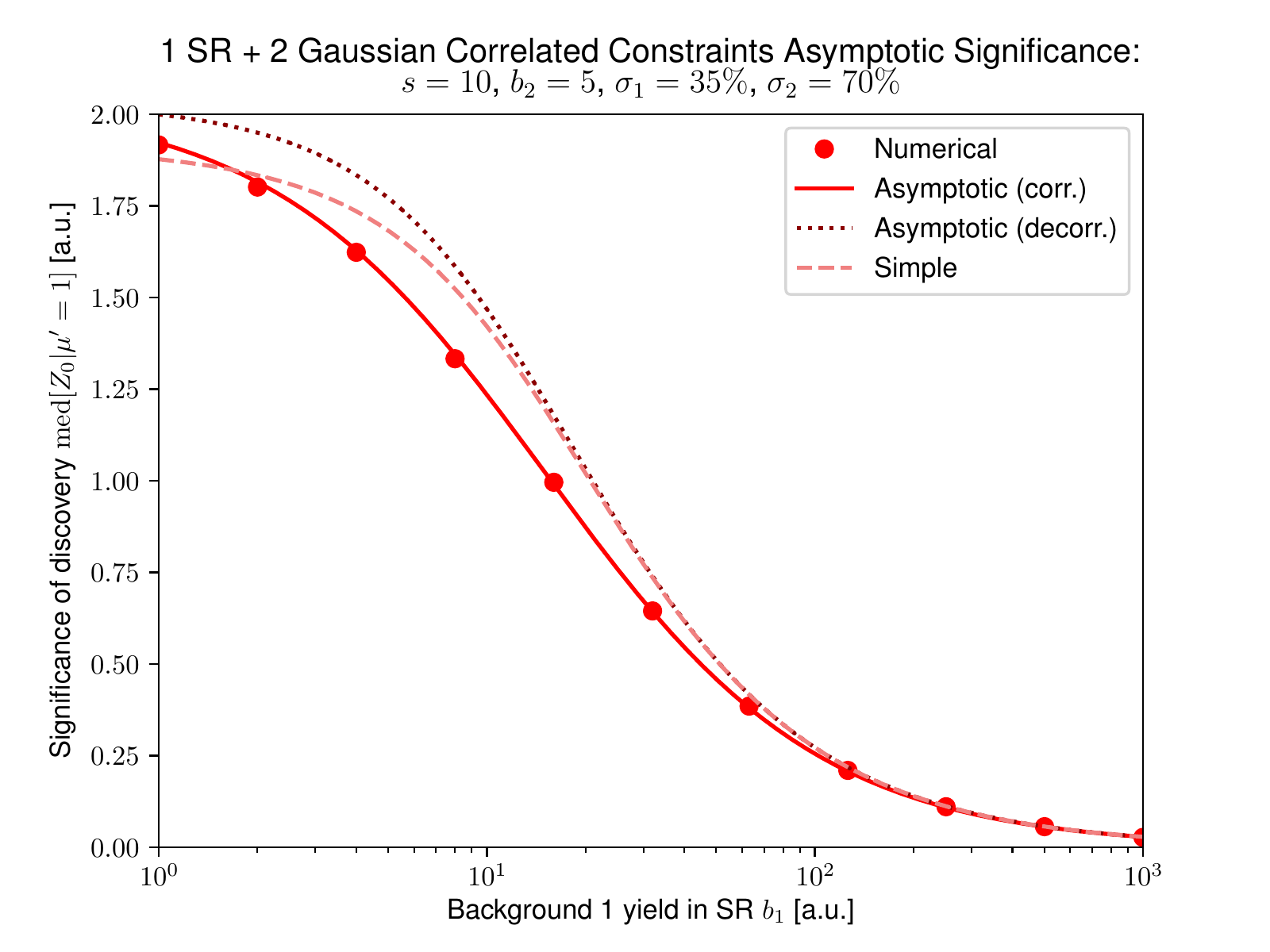}
    \caption{The significance of discovery plotted as a function of the background 1 yield, $b_1$, for Gaussian-constrained measurement described in Section~\ref{sec:gauss_corr}. ``Numerical'' refers to the results calculated using toy-based data (50,000 events for the estimation of $f(t_0|\mu^\prime=0)$ and 50,000 events for the estimation of $f(t_0|\mu^\prime=1)$, per point), ``Asymptotic (corr.)'' refers to the use of Eqs.~\ref{eqn:gauss_thetahh} and \ref{eqn:gauss_z0} (i.e., the correlated case), ``Asymptotic (decorr.)'' refers to the use of Eqs.~\ref{eqn:gauss_decorr_thetahh} and \ref{eqn:gauss_decorr_z0} (i.e., the decorrelated case), and ``Simple'' refers to $s/\sqrt{s+b_1+b_2+\sigma_1^2+\sigma_2^2}$.}
    \label{fig:gauss_z0_corr}
\end{figure}

We may study a situation where these correlations are relevant: two background processes, $b_1$ and $b_2$, and two NPs, $\theta_1$ and $\theta_2$. We assume $\theta_1$ and $\theta_2$ are 75\% correlated and we assume the signal yield $s = 10$ and $b_2 = 5$. We may scan a range of $b_1$ values, but the $s / (b_1 + b_2) \sim 1$ regime is the most interesting. Finally, we assume $b_1$ only responds to $\theta_1$ with response function $R_{b_1}(\theta_1) = 1 + 0.35 \cdot \theta_1$ (i.e., 35\% ``uncertainty'' on the yield $b_1$) and we assume $b_2$ only responds to $\theta_2$ with response function $R_{b_2}(\theta_2) = 1 + 0.7 \cdot \theta_2$ (i.e., 70\% ``uncertainty'' on the yield $b_2$). The sensitivity as a function of $b_1$ for the described measurement scenario is plotted in Fig.~\ref{fig:gauss_z0_corr}.

Below $b_1 = 100$ (i.e., $s / (b_1 + b_2) \lesssim 0.1$), the correlated and decorrelated asymptotic significances begin to diverge. In particular, the decorrelated results (both the asymptotic and ``simple'' approximations) tend to overestimate the sensitivity, which is reduced due to the presence of correlations, as expected. The asymptotic formula which accounts for the 75\% correlation agrees very well with the numerical results, providing excellent validation of the inclusion of such an effect. Above $b_1 = 100$ (i.e., $s / (b_1 + b_2) \gtrsim 0.1$), the total background yield dominates the sensitivity and all three approximations agree.

\subsubsection{Choice of Reponse Function}
\label{sec:response_choice}

In the previous sections, there is freedom in the choice of the response function $R(\theta)$ for interpolating between the up/down response of a background to a particular NP, i.e., subject to the constraints in Eq.~\ref{eqn:response}. For a symmetric up/down response (i.e., $\sigma_{ij}^+ = \sigma_{ij}^-$), a linear response function is the most intuitive (and possibly the only sensible) choice. For an asymmetric up/down response (i.e., $\sigma_{ij}^+ \neq \sigma_{ij}^-$), the situation becomes more complicated. The most intuitive response function is a piecewise linear function:

\begin{equation}
    \label{eqn:linear_response}
    R(\theta) = \left\{
    \begin{array}{ll}
        R_\textrm{up}(\theta)   \,, & \theta \geq 0 \\
        R_\textrm{down}(\theta) \,, & \theta < 0
    \end{array}
    \right. = \left\{
    \begin{array}{ll}
        1 + \sigma_\textrm{up} \cdot \theta \,, & \theta \geq 0 \\
        1 + \sigma_\textrm{down} \cdot \theta \,, & \theta < 0
    \end{array}
    \right. \,, \\
\end{equation}

\noindent where $R_\textrm{up}$ and $R_\textrm{down}$ are the up and down responses, respectively, and $\sigma_\textrm{up}$ and $\sigma_\textrm{down}$ are the up and down relative yield changes, respectively. However, Eq.~\ref{eqn:linear_response} is non-differentiable at $\theta = 0$ -- as Eq.~\ref{eqn:gauss_thetahh} requires derivatives of $R(\theta)$, it is desirable to ensure the response is differentiable for all $\theta$.

One way of ensuring differentiability is to smooth the response function in the vicinity of $\theta=0$ using a weight function $w(\theta)$:

\begin{equation}
    \label{eqn:smooth_linear_response}
    R(\theta) = w(\theta)\cdot R_\textrm{down}(\theta) + (1 - w(\theta))\cdot R_\textrm{up}(\theta) \,,
\end{equation}

\noindent where the weight function is subject to the following constraints:

\begin{equation}
    \label{eqn:weight_req}
    \begin{split}
        w(0) & = \frac{1}{2} \,, \\
        \lim_{\theta\rightarrow -\infty}w(\theta) & = 1 \,, \\
        \lim_{\theta\rightarrow +\infty}w(\theta) & = 0 \,.
    \end{split}
\end{equation}

\noindent The above ensures that the behaviour of the response in the up or down directions is faithfully retained:

\begin{equation}
    \begin{split}
        R(0)  & = R_\textrm{down}(0) = R_\textrm{up}(0) \,, \\
        R(-1) & \approx R_\textrm{down}(-1) \,, \\
        R(+1) & \approx R_\textrm{up}(+1) \,,
    \end{split}
\end{equation}

\noindent as expected. If $w(\theta)$ is smooth, the reponse function is smooth throughout the entire domain.

Any function satisfying Eq.~\ref{eqn:weight_req} may be chosen, examples include:

\begin{equation}
    \label{eqn:weight_functions}
    \setlength{\tabcolsep}{6pt}
    \renewcommand{\arraystretch}{1.5}
    \begin{array}{ll}
        \textrm{Heaviside:} & w(\theta) = 1 - H(\theta) = \left\{
        \begin{array}{ll}
            0 \,,   & \theta > 0 \\
            1/2 \,, & \theta = 0 \\
            1 \,,   & \theta < 0
        \end{array}
        \right. \,, \\
        \textrm{Arctangent:} & w(\theta) = \frac{1}{2}\br{1 - \frac{2}{\pi}\arctan{(\kappa \pi \theta / 2)}} \,, \\
        \textrm{Hyperbolic tangent:} & w(\theta) = \frac{1}{2}\br{1 - \tanh{(\kappa \theta)}} \,, \\
        \textrm{Error function:} & w(\theta) = \frac{1}{2}(1 - \textrm{erf}\,(\kappa \theta)) = \frac{1}{2}\br{1 - \frac{2}{\sqrt{\pi}} \int_0^{\kappa\theta} \exp(-t^2) dt} \,, \\
        \textrm{Sigmoid function:} & w(\theta) = 1 - (1 + \exp(-\kappa\theta))^{-1} \,. \\
    \end{array}
\end{equation}

\noindent The parameter $\kappa$ tunes how sharply the transition is between the up/down responses at $\theta=0$. N.B.: the Heaviside weighting function is identical to the non-differentiable case, Eq.~\ref{eqn:linear_response}.

\begin{figure}[htbp]
    \centering
    \begin{subfigure}{.49\textwidth}
        \centering
        \includegraphics[width=1.1\linewidth]{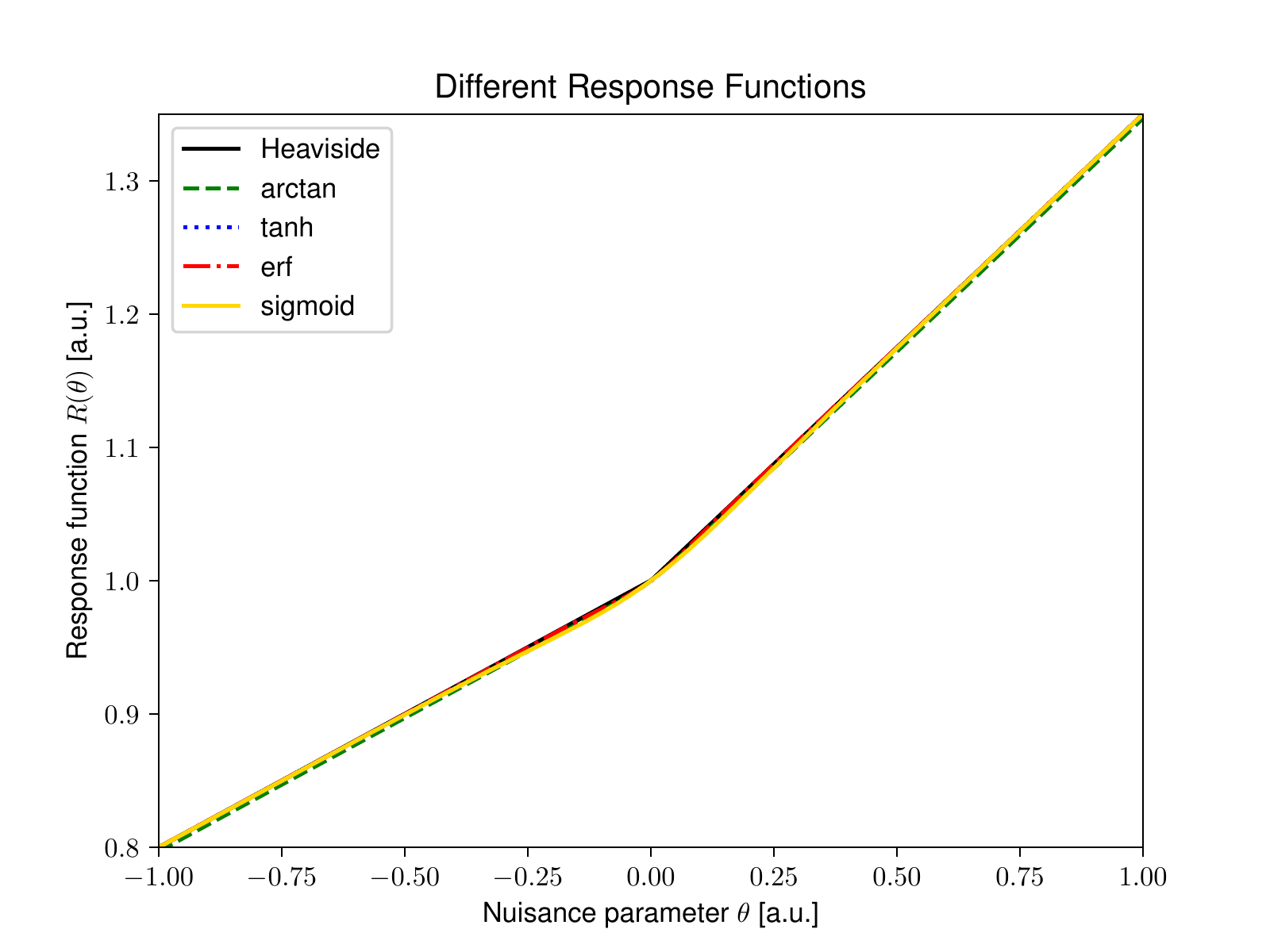}
        \caption{Weighting functions.}
    \end{subfigure}
    \begin{subfigure}{.49\textwidth}
        \centering
        \includegraphics[width=1.1\linewidth]{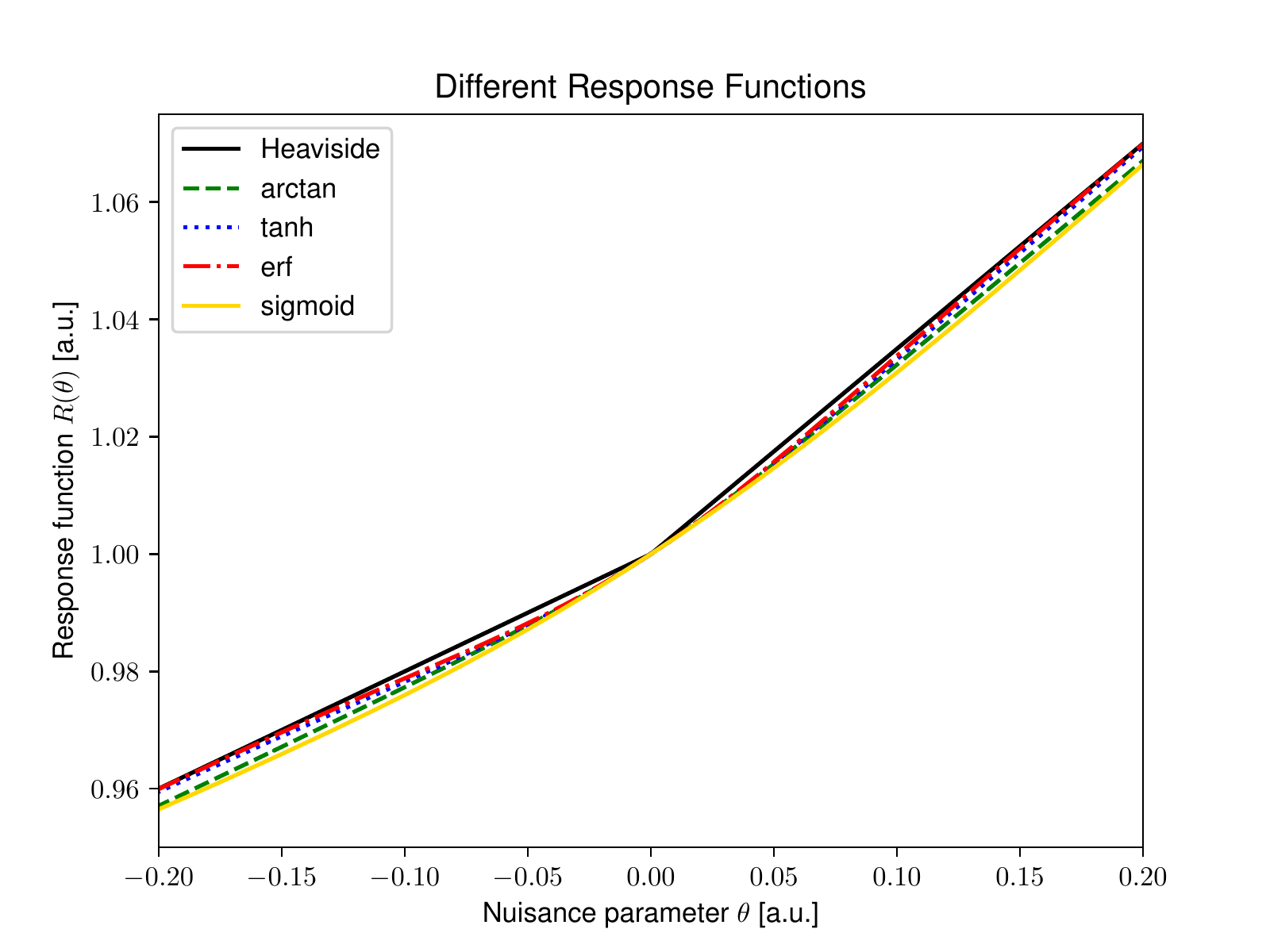}
        \caption{Weighting functions, zoomed.}
    \end{subfigure} \\
    \centering
    \begin{subfigure}{.49\textwidth}
        \centering
        \includegraphics[width=1.1\linewidth]{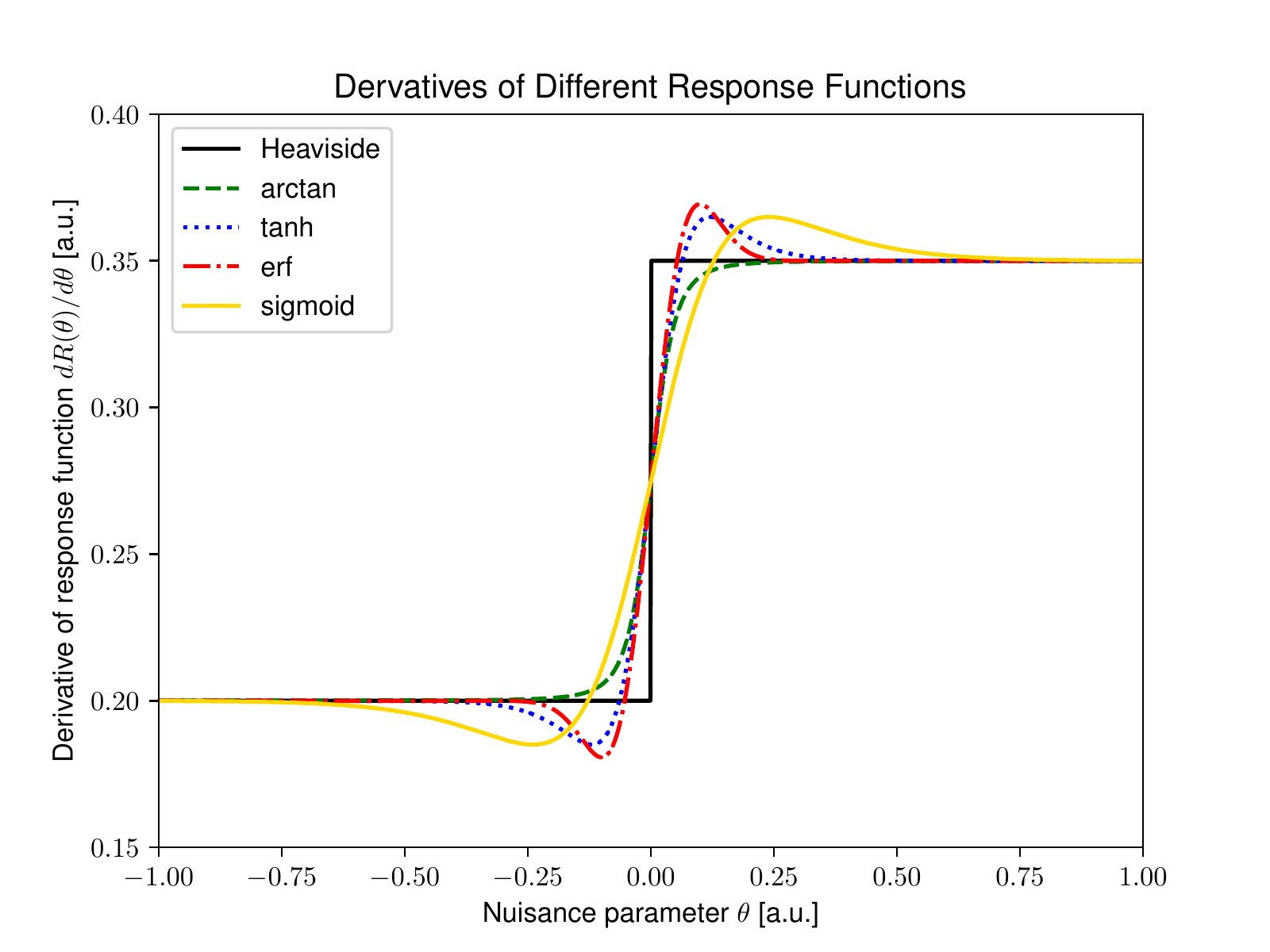}
        \caption{First derivatives of weighting functions.}
    \end{subfigure}
    \caption{Equation~\ref{eqn:smooth_linear_response} is plotted for the different choices of the weighting function, described in Eq.~\ref{eqn:weight_functions}, in (a) and zoomed in on $\theta=0$ in (b). The first derivative of Eq.~\ref{eqn:smooth_linear_response} is plotted for the different choices of the weighting function in (c). In all plots, $\sigma_\textrm{down} = 20\%$ and $\sigma_\textrm{up} = 35\%$. The parameter $\kappa$ in the weighting functions is 10.}
    \label{fig:response_functions}
\end{figure}

In Fig.~\ref{fig:response_functions}, the effect of the choice of weighting function on the response function, Eq.~\ref{eqn:smooth_linear_response}, and its first derivative are plotted. The hyperbolic tangent and error functions result in responses which converge the fastest to that using the Heaviside weighting function, while the arctangent and sigmoid functions result in responses which converge much more slowly. In contrast, the hyperbolic, error, and sigmoid functions result in large overshoots for the first deratives of these response functions, while the arctangent function results in a response whose first derivative converges the fastest to that using the Heaviside weighting function. With these remarks in mind, it may be concluded that the sigmoid weighting function is the poorest choice of weighting function for smoothing the behaviour of an asymmetric response about $\theta=0$. All choices result in a response which is differentiable at $\theta=0$, as desired.

\begin{figure}[htbp]
    \centering
    \begin{subfigure}{.45\textwidth}
        \centering
        \includegraphics[width=1.1\linewidth]{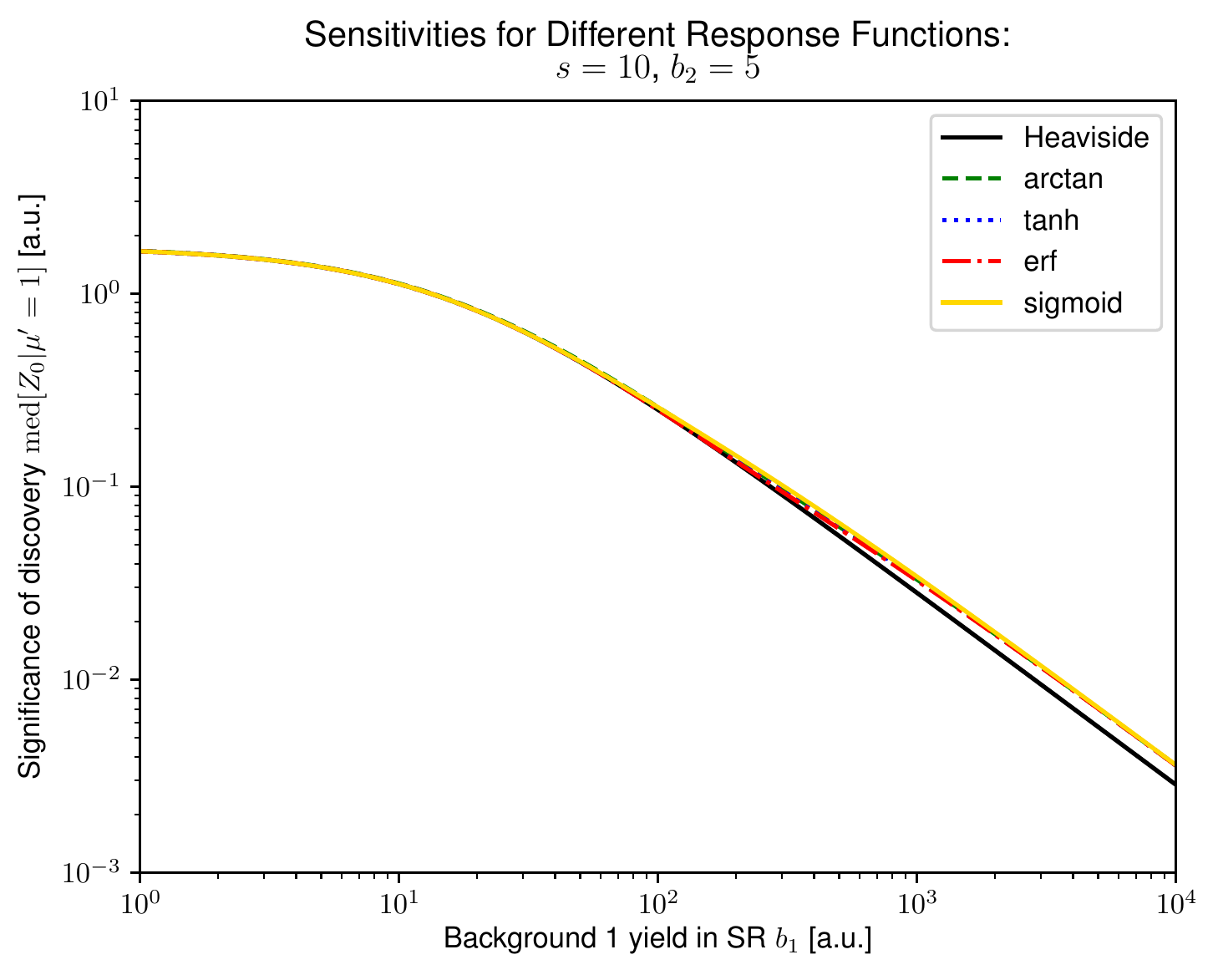}
        \caption{$(s, b_2) = (10, 5)$.}
        \label{fig:gauss_z0_response_choice_low_b2}
    \end{subfigure}
    \hspace{.03\textwidth}
    \begin{subfigure}{.45\textwidth}
        \centering
        \includegraphics[width=1.1\linewidth]{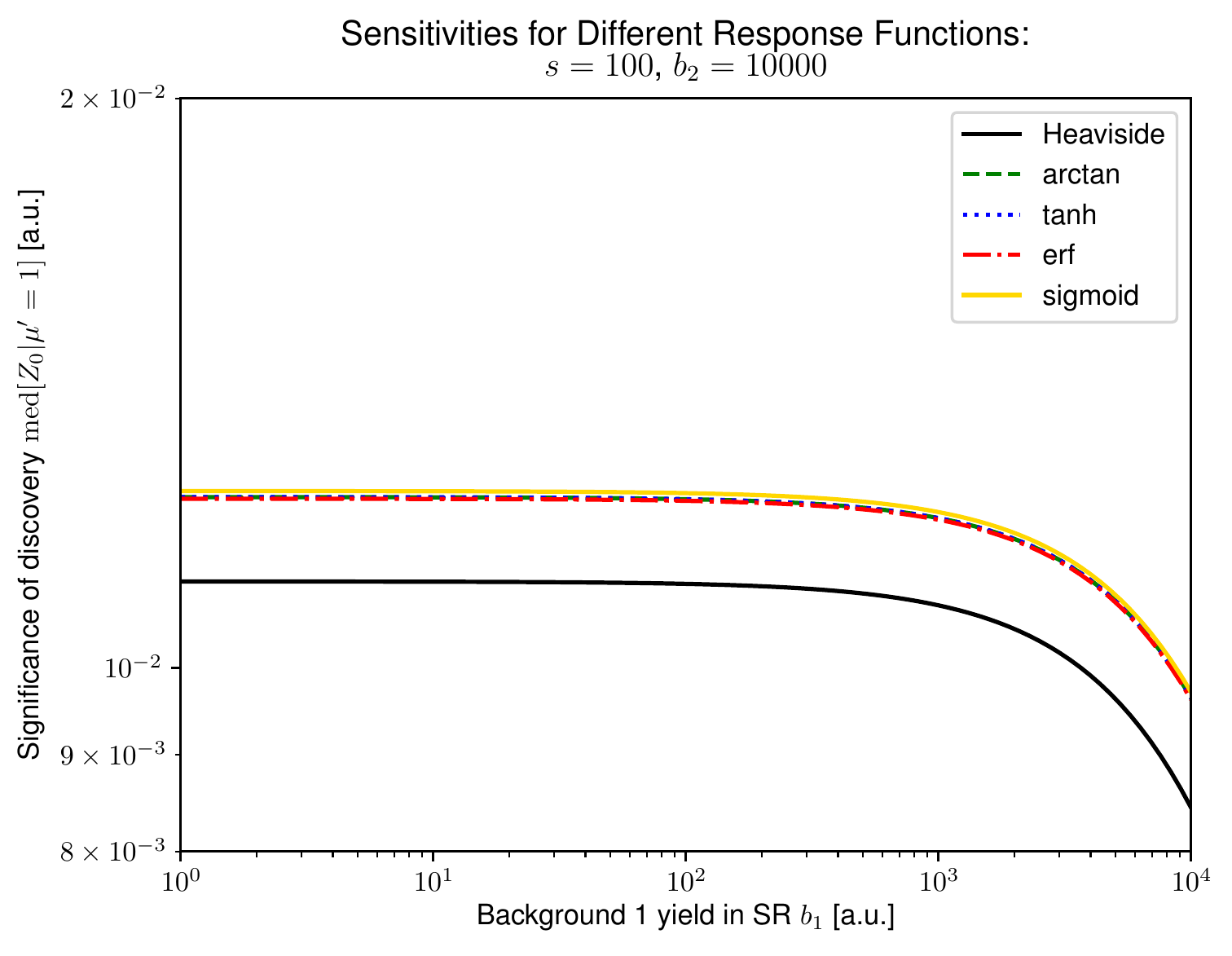}
        \caption{$(s, b_2) = (100, 10000)$.}
        \label{fig:gauss_z0_response_choice_high_b2}
    \end{subfigure}
    \caption{The significance of discovery plotted as a function of the background 1 yield, $b_1$, for different choices of response function (read: weighting function for interpolating about $\tvec = \vec{0}$) for the Gaussian-constrained measurement described in Section~\ref{sec:response_choice}. Two different choices of signal yield, $s$, and the background 2 yield, $b_2$, are tested as well in (a) and (b).}
    \label{fig:gauss_z0_response_choice}
\end{figure}

We can also look at the effect of the choice of response function on the sensitivity calculated using asymptotic approximations. We consider a measurement scenario with a signal yield $s$, two background processes, $b_1$ and $b_2$, and two NPs, $\theta_1$ and $\theta_2$. We assume $\theta_1$ and $\theta_2$ are decorrelated. Additionally, we assume $b_1$ only responds to $\theta_1$ with up and down uncertainties of 35\% and 20\%, respectively, and we assume $b_2$ only responds to $\theta_2$ with up and down uncertainties of 90\% and 70\%, respectively. Using the function \texttt{asymptotic\_formulae.GaussZ0} from Ref.~\cite{Basso2021} to solve Eq.~\ref{eqn:gauss_thetahh} and evaluate Eq.~\ref{eqn:gauss_z0}, we have plotted the sensitivity as a function of the background yield -- this is shown in Fig.~\ref{fig:gauss_z0_response_choice}.

From Fig.~\ref{fig:gauss_z0_response_choice_low_b2}, we see the choice of response function as \textit{no effect} on the estimated sensitivity when considering the scale of the plot between $b_1 = 1$ and $b_1 = 10,000$. Deviations between the Heaviside weighting function and the others are only present in the high background regime -- this is demonstrated in Fig.~\ref{fig:gauss_z0_response_choice_high_b2}. While a difference between the Heaviside weighting function and the others is relevant when considering the scale of the plot, the sensitivity is so low in this regime that the effect is of no practical importance. It may be concluded that the user can safely choose any of the described response functions for smoothing the behaviour near $\theta = 0$ (if differentiability at that point is important) without incurring a systematic effect on the estimated sensitivity.

\subsubsection{CPU Performance}
\label{sec:cpu}

It is also interesting to compare the CPU performance of the asymptotic formulae to the toy-based approach. We consider a complicated likelihood scenario where we have a single SR bin and two background processes in that bin: $s = 10$, $b_1 = 10$, and $b_2 = 5$. We also assume a single NP $\theta_1$ tunes the response of $b_1$ and a single NP $\theta_2$ tunes the response of $b_2$ and that these two NPs have a 75\% correlation. We assume the response functions of $b_1$ and $b_2$ are $1 + 0.35 \cdot \theta_1$ (i.e., 35\% ``uncertainty'' on $b_1$) and $1 + 0.70 \cdot \theta_2$ (i.e., 70\% ``uncertainty'' on $b_2$), respectively. A derivation of the asymptotic formula covering the described likelihood does not exist in the literature outside of this paper. As a result, the only recourse available is using toy-based data, making the described likelihood a good example of the benefit of using the asymptotic approximations of Wilks and Wald over throwing toy data.

To estimate the sensitivity using asymptotic approximations, \texttt{asymptotic\_formulae.GaussZ0} from Ref.~\cite{Basso2021} to solve Eq.~\ref{eqn:gauss_thetahh} and evaluate Eq.~\ref{eqn:gauss_z0}. To estimate the sensitivity using toy-based data, the function \texttt{asymptotic\_formulae.GaussZ0\_MC} from Ref.~\cite{Basso2021} is used. The author does not profess to have written the most efficient code possible for either function -- this study is mostly to demonstrate roughly the order-of-magnitude CPU time for each. If relevant, the author's machine uses an Intel Core i7-6500U CPU (4x 2.50GHz) and 12 GB DDR3 RAM. The CPU time is monitored before and after the relevant function calls.

\begin{figure}[htbp]
    \centering
    \includegraphics[width=0.9\textwidth]{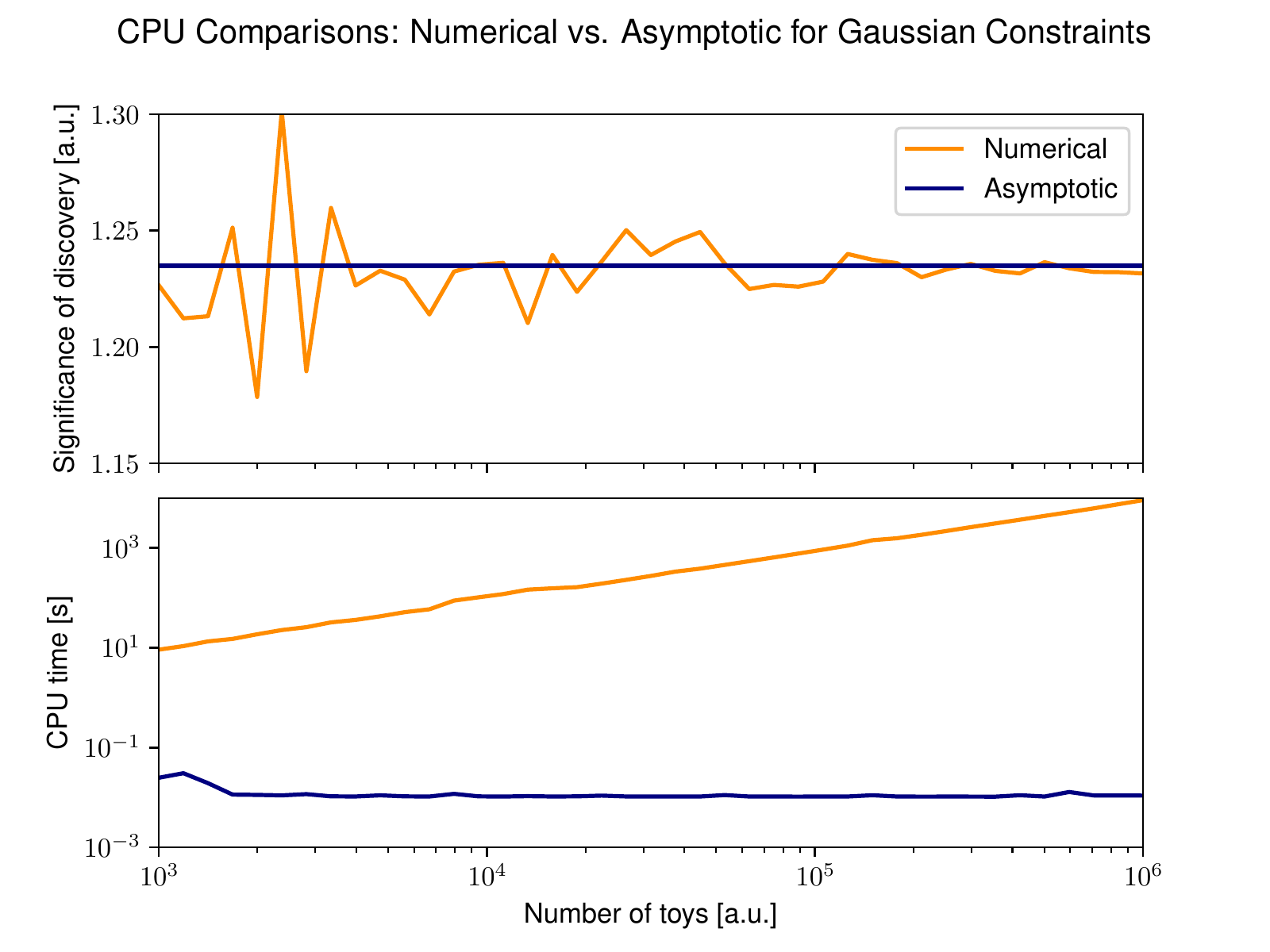}
    \caption{The significance of discovery (top) and CPU performance (bottom) for both asymptotic and toy-based (``numerical'') approaches for Gaussian-constrained measurement described in Section~\ref{sec:cpu}. Forty-one points are evaluated per curve. N.B.: the number of toys only affects the toy-based approach, but the asymptotic formula is re-run and re-timed at each point.}
    \label{fig:cpu}
\end{figure}

Figure~\ref{fig:cpu} shows the estimated sensitivity of the measurement for both the numerical and asymptotic approaches as a function of the number of toys used in the numerical approach. N.B.: the asymptotic formula is not affected by the number of toys, but it is evaluated and timed at each step -- additionally, the asymptotic formula's result is not expected to change at each step. As expected, the numerical approach converges to the asymptotic formula's result as the number of toys becomes large. The time required by the numerical approach increases linearly from $\mathcal{O}(10^1)$~s to $\mathcal{O}(10^3)$~s as the number of toys increases from 1,000 to 1,000,000. In contrast, the time required by using the asymptotic approximations (accounting for the time required to solve the system of equations yielding $\thhvec$) is $\mathcal{O}(10^{-2})$~s. To obtain reasonably stable (i.e., fluctuations in $Z_0$ less than 1\%) toy-based results, roughly $\mathcal{O}(10,000)$ toys or more are necessary. Thus, for most practical use cases, one can expect the asymptotic results to be 3-4 orders of magnitude faster than the equivalent toy-based approach.

\FloatBarrier

\FloatBarrier

\section{Summary}
\label{summary}
We have presented a collection of derivations of generalized formulae for estimating the median significance of discovery in the asymptotic limit for various measurement models in HEP. The formulae have been verified to agree with numerical results using toy-based data in select cases; other times, they are shown to reduce to known formulae in simpler cases derived by similar means. In the low $s/b$ regime, simpler versions of the asymptotic formulae based on $s/\sqrt{s+b}$ do just as well as the more accurate formulae derived in this document, agreeing with the conclusion of Ref.~\cite{Cowan2011} that these simple formulae work for $s/b \ll 1$. In the $s/b \sim 1$ regime, we show that the formulae derived in this document agree well with numerical results whereas the simpler versions fail. A summary of the different measurement scenarios considered in this paper (or elsewhere and rederived in this paper) as well as the relevant asymptotic signficance formulae are detailed in Table~\ref{tab:summary}.

Possible extensions to this work could include deriving the significance for $N$ SRs + $M$ CRs in Section~\ref{sec:NSRMCR} when the CRs are assumed to pure in each background (i.e., diagonal $\boldsymbol{\tau}^i \,\forall\, i = 1,\ldots,N$), generalizing the type of constraint in Section~\ref{sec:gauss} (e.g., Gaussian, log-normal, etc.), generalizing the NPs in Section~\ref{sec:gauss} to also act on signal, and generalizing the formulae in Section~\ref{sec:gauss} to include an arbitrary number of SRs in addition to an arbitrary number of backgrounds and NPs.

\begin{table}[htbp]
    \centering
    \caption{A summary of the measurement scenarios considered in this paper and the asymptotic formuale for the significance of discovery derived in each case. The relevant sections or references (in the cases where the formulae have been derived elsewhere and are reproduced here) are given as well as the pertinent equations and/or Python functions for applying them.}
    \label{tab:summary}
    \resizebox{\textwidth}{!}{
        \begin{tabular}{lll}
            \toprule\toprule
            Measurement & Reference or section & Formula or function \\
            \midrule
            \multirow{2}{*}{1 SR + $N$ CRs, arbitrary $\tau$}                                 & \multirow{2}{*}{Section~\ref{sec:1SRNCR_general}}       & Eq.~\ref{eqn:ncr_z0} using Eq.~\ref{eqn:ncr_bhh} \\
                                                                                              &                                                         & \,\,\,\, or \texttt{asymptotic\_formulae.nCRZ0} (Ref.~\cite{Basso2021}) \\
            1 SR + 1 CR                                                                       & Ref.~\cite{Cowan2011,Buttinger2019}                     & Eq.~\ref{eqn:1cr_simple_z0} \\
            \multirow{2}{*}{1 SR + 2 CRs, diagonal $\tau$}                                    & \multirow{2}{*}{Section~\ref{sec:1SRNCR_Neq2_diag_tau}} & Eq.~\ref{eqn:ncr_z0_diag} using Eqs.~\ref{eqn:2cr_b2} and \ref{eqn:2cr_b1} \\
                                                                                              &                                                         & \,\,\,\, or \texttt{make\_paper\_plots.GaussZ0\_DecorrConstAndNeqMeq2} (Ref~\cite{Basso2021}) \\
            \midrule
            \multirow{2}{*}{$N$ SRs + 1 CR, correlated $\mu$}                                 & \multirow{2}{*}{Section~\ref{sec:NSR1CR}}               & Eq.~\ref{eqn:nsr_z0} using Eq.~\ref{eqn:nsr_bhh} \\
                                                                                              &                                                         & \,\,\,\, or \texttt{asymptotic\_formulae.nSRZ0} (Ref.~\cite{Basso2021}) \\
            \midrule
            $N$ SRs + $M$ CRs, correlated $\mu$                                               & Section~\ref{sec:NSRMCR}                                & Eq.~\ref{eqn:nsr_mcr_z0} using Eq.~\ref{eqn:nsr_mcr_bhh_eqns} \\
            \midrule
            1 SR + $N$ backgrounds                                                            & \multirow{2}{*}{Section~\ref{sec:gauss_general}}        & Eq.~\ref{eqn:gauss_z0} using Eq.~\ref{eqn:gauss_thetahh} \\
            \,\,\,\, with $M$ correlated Gaussian constraints                                 &                                                         & \,\,\,\, or \texttt{asymptotic\_formulae.GaussZ0} (Ref~\cite{Basso2021}) \\
            1 SR + $N$ backgrounds                                                            & \multirow{2}{*}{Section~\ref{sec:gauss_decorr}}         & Eq.~\ref{eqn:gauss_decorr_z0} using Eq.~\ref{eqn:gauss_decorr_bhh} \\
            \,\,\,\, with $N$ decorrelated Gaussian constraints                               &                                                         & \,\,\,\, or \texttt{asymptotic\_formulae.GaussZ0} (Ref~\cite{Basso2021}) \\
            1 SR + 1 background with 1 Gaussian constraint                                    & Ref.~\cite{Buttinger2019}                               & Eq.~\ref{eqn:gauss_simple_z0} using Eq.~\ref{eqn:gauss_simple_bhh} \\
            \bottomrule\bottomrule
        \end{tabular}
    }
\end{table}

\FloatBarrier

\clearpage
\bibliographystyle{spphys} 
\bibliography{Basso_Generalized_Asymptotic_Formulae} 

\clearpage
\begin{appendices}

\section{Solution to $\frac{A}{\sum_j x_j} + B_ix_i + C_i=0 \,,\,i=1,\ldots,N$}
\label{sec:appA}
\numberwithin{equation}{section}
\setcounter{equation}{0}
\renewcommand{\theequation}{\thesection\arabic{equation}}

We consider the solutions to the system of equations:

\begin{equation}
    \label{eqn:appA_system}
    \cbr{\frac{A}{\sum_{j=1}^{N}x_{j}} + B_i x_i + C_i = 0 \,;\, i = 1,\ldots,N} \,,
\end{equation}

\noindent where $A$ is a constant and $B_i$ and $C_i$ are equation-dependent constants, analogous to Eq.~\ref{eqn:gauss_decorr_bhh}. Subtracting the $j$-th equation from the $i$-th equation:

\begin{equation}
    B_i x_i + C_i - B_j x_j - C_j = 0 \Leftrightarrow x_j = \frac{B_i x_i + (C_i - C_j)}{B_j} \,,
\end{equation}

\noindent which can then be inserted into the $i$-th equation of Eq.~\ref{eqn:appA_system} to yield an expression entirely in terms of $x_i$:

\begin{equation}
    \label{eqn:appA_xi}
    \begin{split}
        & A + B_i x_i \sum_{j=1}^N x_j + C_i \sum_{j=1}^N x_j = 0\\
        & \Leftrightarrow A + B_i x_i \sum_{j=1}^N\frac{B_i x_i + (C_i - C_j)}{B_j} + C_i \sum_{j=1}^N\frac{B_i x_i + (C_i - C_j)}{B_j} = 0 \\
        & \Leftrightarrow \br{B_i^2\sum_{j=1}^N\frac{1}{B_j}}x_i^2 + \br{B_i \sum_{j=1}^N\frac{2 C_i-C_j}{B_j}} x_i + \br{A + C_i\sum_{j=1}^N\frac{C_i - C_j}{B_j}} = 0 \,,
    \end{split}
\end{equation}

\noindent which may be solved using the quadratic equation, selecting the real and positive root for $x_i$ (or the answer which statisfies the system of equations as the physical solution as $x_i \leftrightarrow \bhh_i$). From Eq.~\ref{eqn:gauss_decorr_bhh}, we identify $A=n$, $B_i=-1/\sigma_i^2$, and $C_i=b_i/\sigma_i^2-1$. Thus in Eq.~\ref{eqn:appA_xi}, we know:

\begin{equation}
    B_i^2\sum_{j=1}^N \frac{1}{B_j} = \frac{-1}{\sigma_i^4}\sum_{j=1}^N\sigma_j^2 < 0 \,,
\end{equation}

\noindent as $\sigma_i > 0 \,\forall\, i$. From Eq.~\ref{eqn:appA_xi}, we identify the quadratic equation:

\begin{equation}
    -|\tilde{A}_i|x^2+\tilde{B}_ix+\tilde{C}_i=0 \,,
\end{equation}

\noindent where:

\begin{equation}
    \label{eqn:appA_tilde}
    \begin{split}
        |\tilde{A}_i| & = \left|B_i^2\sum_{j=1}^N \frac{1}{B_j}\right| = \frac{1}{\sigma_i^4}\sum_{j=1}^N\sigma_j^2 \,, \\
        \tilde{B}_i & = B_i \sum_{j=1}^N\frac{2 C_i-C_j}{B_j} = \frac{1}{\sigma_i^2}\sum_{j=1}^N \sigma_j^2\cdot\br{\frac{2b_i}{\sigma_i^2} - \frac{b_j}{\sigma_j^2} - 1} \,, \\
        \tilde{C}_i & = A + C_i\sum_{j=1}^N\frac{C_i - C_j}{B_j} = n + \br{1 - \frac{b_i}{\sigma_i^2}} \cdot \sum_{j=1}^N \sigma_j^2 \cdot \br{\frac{b_i}{\sigma_i^2} - \frac{b_j}{\sigma_j^2}} \,,
    \end{split}
\end{equation}

\noindent which has the solutions:

\begin{equation}
    \label{eqn:appA_solX}
    x_i=\frac{\tilde{B}_i \pm \sqrt{\tilde{B}_i^2+4|\tilde{A}_i|\tilde{C}_i}}{2|\tilde{A}_i|} \,.
\end{equation}

Using Eq.~\ref{eqn:appA_tilde}, it can be shown that:

\begin{equation}
    \label{eqn:appA_Btilde}
    \tilde{B}_i = \frac{\tilde{C}_i - n}{\sigma_i^2 - b_i} - (\sigma_i^2 - b_i)\cdot|\tilde{A}_i| \,,
\end{equation}

\noindent and therefore:

\begin{equation}
    \begin{split}
        \tilde{B}_i^2+4|\tilde{A}_i|\tilde{C}_i & = \br{\frac{\tilde{C}_i - n}{\sigma_i^2 - b_i}}^2 + (\sigma_i^2 - b_i)^2\cdot|\tilde{A}_i|^2 - 2\cdot(\tilde{C}_i-n)\cdot|\tilde{A}_i| + 4|\tilde{A}_i|\tilde{C}_i \\
        & = \br{\frac{\tilde{C}_i - n}{\sigma_i^2 - b_i}}^2 + (\sigma_i^2 - b_i)^2\cdot|\tilde{A}_i|^2 + 2\cdot(\tilde{C}_i-n)\cdot|\tilde{A}_i| + 4n|\tilde{A}_i| \\
        & = \br{\frac{\tilde{C}_i - n}{\sigma_i^2 - b_i} + (\sigma_i^2 - b_i)\cdot|\tilde{A}_i|}^2 + 4n|\tilde{A}_i| \,,
    \end{split}
\end{equation}

\noindent which is always positive, so we are \textit{always} guaranteed a real root in Eq.~\ref{eqn:appA_solX}. We consider the difference:

\begin{equation}
    \br{\frac{\tilde{C}_i - n}{\sigma_i^2 - b_i} + (\sigma_i^2 - b_i)\cdot|\tilde{A}_i|} - \br{\frac{\tilde{C}_i - n}{\sigma_i^2 - b_i} - (\sigma_i^2 - b_i)\cdot|\tilde{A}_i|} = 2 \cdot (\sigma_i^2 - b_i)\cdot|\tilde{A}_i| \,.
\end{equation}

\noindent If $\sigma_i^2 > b_i$, the above is \textit{positive}: this means the argument of our square root will \textit{always} be larger than $\tilde{B}_i$ and we \textit{must} select the positive sign solution for the $i$-th equation for a physical solution. By Eq.~\ref{eqn:appA_Btilde}, we know:

\begin{equation}
    \tilde{B}_i = \frac{1}{\sigma_i^2} \cdot \sum_{j=1}^N \sigma_j^2 \cdot \br{\frac{b_i}{\sigma_i^2} - \frac{b_j}{\sigma_j^2}} +  (b_i - \sigma_i^2)\cdot|\tilde{A}_i| \,.
\end{equation}

\noindent If $b_i > \sigma_i^2$, the second term in the above is positive. And if $\sum_{j=1}^N \sigma_j^2 \cdot \br{b_i/\sigma_i^2 - b_j/\sigma_j^2} < 0$ and $|\sum_{j=1}^N \sigma_j^2 \cdot \br{b_i/\sigma_i^2 - b_j/\sigma_j^2}| > (b_i - \sigma_i^2)\cdot|\tilde{A}_i|$, then we \textit{must} select the positive sign solution for a physical solution. In the case where $\sum_{j=1}^N \sigma_j^2 \cdot \br{b_i/\sigma_i^2 - b_j/\sigma_j^2} > 0$, we must test both solutions for the $i$-th equation amongst all other solutions and pick the one which satisfies our system of equations and yields positive solutions $x_i > 0\,\forall\,i=1,\ldots,N$. In this way, we have not precisely solved our system of equations, but we have set an upper limit of $N\times N$ solutions to be explored before the physical solution is found.

\FloatBarrier

\end{appendices}

\end{document}